\documentclass[apj]{emulateapj}
\usepackage{graphicx,floatrow,amsmath,gensymb,rotating}
\usepackage{subfigure}
\usepackage{natbib}
\usepackage{xcolor}

\newcommand{\msunh}{\>h^{-1}\rm M_\odot}

\begin{document}

\title{THE MORPHOLOGICAL TRANSFORMATION AND THE QUENCHING OF GALAXIES}
\author{
Chenxu~Liu\altaffilmark{1,2,3},
Lei~Hao\altaffilmark{1},
Huiyuan~Wang\altaffilmark{4,5},
Xiaohu~Yang\altaffilmark{6,7}
}
\email{haol@shao.ac.cn}

\altaffiltext{1}{Key Laboratory for Research in Galaxies and Cosmology of Chinese Academy of Sciences, Shanghai Astronomical Observatory, 80 Nandan Road, Shanghai 200030, China}
\altaffiltext{2}{University of Chinese Academy of Sciences, Beijing 100049, China}
\altaffiltext{3}{Department of Astronomy, University of Texas, Austin, TX 78712, USA}
\altaffiltext{4}{Key Laboratory for Research in Galaxies and Cosmology, Department of Astronomy, University of Science and Technology of China, Hefei, Anhui 230026, China}
\altaffiltext{5}{School of Astronomy and Space Science, University of Science and Technology of China, Hefei 230026, China}
\altaffiltext{6}{Department of Astronomy, and Tsung-Dao Lee Institute, Shanghai Jiao Tong University, Shanghai 200240, China}
\altaffiltext{7}{Shanghai Key Laboratory for Particle Physics and Cosmology, Shanghai Jiao Tong University, Shanghai 200240, China}

\begin{abstract}
We study the morphological transformation from late types to early types and the quenching of galaxies with the seventh Data Release (DR7) of the Sloan Digital Sky Survey (SDSS). Both early type galaxies and late type galaxies are found to have bimodal distributions on the star formation rate versus stellar mass diagram ($\lg SFR - \lg M_*$). We therefore classify them into four types: the star-forming early types (sEs), the quenched early types (qEs), the star-forming late types (sLs) and the quenched late types (qLs). We checked many parameters on various environmental scales for their potential effects on the quenching rates of late types and early types, as well as the early type fractions among star-forming galaxies and those among quenched galaxies. These parameters include: the stellar mass $M_*$, and the halo mass $M_{halo}$; the small-scale environmental parameters, such as the halo centric radius $R_p/r_{180}$ and 
the third nearest neighbor distances ($d_{3nn}$); the large-scale environmental parameters, specifically whether they are located in clusters, filaments, sheets, or voids. We found that 
the morphological transformation is mainly regulated by the stellar mass. Quenching is mainly driven by the stellar mass for more massive galaxies and by the halo mass for galaxies with smaller stellar masses. 
In addition, we see an overall stronger halo quenching effect in early type galaxies, which might be attributed to their lacking of cold gas or earlier accretion into the massive host halos. 
\end{abstract}


\section{Introduction}
\label{sec_intro}

Galaxies are found to be clustered into two populations on the color-magnitude diagram (CMD), or the $\lg SFR -\lg M_*$ diagram in the imaging and spectroscopic surveys  \citep[e.g.][]{Strateva2001,Baldry2004,Bell2004,Blanton2005,Faber2007,Wetzel2012}. These two populations are often referred as the blue cloud and the red sequence. The blue cloud galaxies are less massive, less luminous, actively forming stars, and of younger stellar populations. The red sequence galaxies are more massive, more luminous, with little or no star formation, and of older stellar populations. As the cold gas reservoir within galaxies is used up or deprived by various mechanisms, the star formation will shut down and galaxies will evolve from the blue cloud to the red sequence \citep[e.g.][]{Faber2007}.

The morphologies of galaxies are also found to have two distinct populations: the early-type galaxies and the late-type galaxies \citep{Hubble1936}. Late-type galaxies have disc-dominated morphologies with spiral arms in their disks. Early-type galaxies are ellipiticals or bulge-dominated S0s with no arm features.

The morphology and the star formation state of galaxies are usually considered to have a degeneracy. The red sequence galaxies usually have early-type morphologies and the blue cloud galaxies tend to be late type spirals. 
An explanation for this degeneracy is when the star formation of galaxies got quenched and galaxies evolve from the blue cloud to the red sequence, the morphology may be transformed simultaneously. However, there are at least two types of galaxies that show the situation may be far more complicated than assumed: one is
the star-forming early type galaxies, sEs, which have early-type morphologies, but are still actively forming stars \citep[e.g.][]{Bamford2009,Kannappan2009,HC2010,George2017}; the other is the quenched late types, qLs, which have late-type morphologies, but are quenched and red  \citep[e.g.][]{vdBergh1976,Bekki2002,Goto2003,Ishigaki2007,Bamford2009,Wolf2009,Skibba2009,Bundy2010,Masters2010,Rowlands2012,F-M2018}. 

sEs are mainly considered to be formed in major merger events because of their high asymmetry index. The star formation is not fully shut down right after the merger events, while bulges are already built and disks are destroyed. In this scenario, morphology transformation happens before quenching finishes. sEs could alternatively be the results of the rejuvenation of quenched early types (qEs) by external sources \citep[e.g.][and the references therein]{Kim2018,Kannappan2009}. Environments could also be an explanation for their formation mechanism. We will talk about the environments later in the following paragraphs.
The formation mechanisms of qLs are thought to be different at different stellar masses. Less massive qLs are mostly found in clusters and were ram-pressure stripping and/or strangulation quenched. For massive qLs, both internal and external processes play roles in their quenching \citep[e.g.][and the references therein]{F-M2018}. In both scenarios, quenching happens before morphology changing.

It is therefore interesting to look closer at the quenching of star formation and the transformation of the morphologies separately, so that our results will not be affected by the SFR-morphology degeneracy. 
There are many mechanisms that can possibly affect both of these two important changes in the life time of galaxies. One important factor is the environment.
A color-density relation was found that dense environments are mostly populated by red passive galaxies and underdense environments are favored by the blue star-forming galaxies  \citep[e.g.][]{Lewis2002,Gomez2003,Kauffmann2004,Rojas2005,Weinmann2006,Bamford2008,Constantin2008,Tinker2008,SkibbaSheth2009,Hoyle2012,Liu2015,Moorman2016}.
A morphology-density relation was also detected that the fraction of elliptical galaxies increases and that of spiral galaxies decreases with increasing galaxy density  \citep[e.g.][]{Dressler1980,Postman1984,Norberg2002,Goto2003,Blanton2005md,Wolf2007,Ball2008}. 

These relationships suggest that environmental effects may be an important parameter in regulating the star-formation quenching or morphology transformation, or both at the same time. Possible mechanisms that are environmental-related include violent merger events \citep[e.g.][]{Toomre1972}, gentle interactions \citep[e.g.][]{Moore1996,Moore1999,Fujita1998}, ram pressure stripping \citep[e.g.][]{Gunn1972,Farouki1980,Fujita1999,Quilis2000}, or the gradually halted supply of cold gas in clusters (the so-called ``strangulation'' or ``starvation'') \citep[e.g.][]{Larson1980,Bekki2002,Peng2015}, etc. \cite{Schawinski2014} studied the morphology related star formation histories of galaxies with galaxy zoo and found two evolutionary pathways of quenching: the rapid quenching of blue early types, possibly driven by major mergers; and the slow quenching of late types, quenched through secular evolution. \cite{Smethurst2015} further found a third pathway of quenching -- the intermediate morphology dominated intermediate quenching, which may be caused by weaker interactions. However, it is also suggested that the environment may not be the most fundamental parameter in the quenching process, which seems to be more correlated with the stellar mass or halo mass \citep[e.g.][]{Liu2015,WangE2018,WangH2018}.

We checked various parameters on various scales to study whether quenching and morphological transformation are fundamentally driven by the same parameter(s) and what they are. From small to large scale, we studied the stellar mass $M_*$ tracing the galaxy scale, the halo mass $M_{halo}$ and the halo centric radius $R_p/r_{180}$ representing the halo scale, the  third nearest neighbor distances ($d_{3nn}$) as the indicator of the small-scale environments, and the large-scale environments (clusters, filaments, sheets, and voids). Section \ref{sec_data} introduces the sample and the derived properties used in our analysis, Section \ref{sec_results} presents our main results, Section \ref{sec_discuss} discusses the implications of our results, and Section \ref{sec_conclusions} summarizes our main conclusions.


\section{DATA}
\label{sec_data}

\subsection{The galaxy Sample}
\label{subsec_parentsample}

Our parent galaxy sample is taken from \cite{WangH2018}. They select all galaxies in the main galaxy sample of the New York University Value-added Galaxy Catalog (NYU-VAGC; \cite{Blanton2005}) of the SDSS DR7 \citep{aba09}, with (1) r-band apparent magnitudes $r\le$ 17.72; (2) redshift completeness $C\ge$0.7; (3) within the reconstruction region of ELUCID simulation (see \cite{Wang2016} for details); (4) the stellar mass $M_*>10^{8.5}\ h^{-2} M_{\sun}$. The selection criteria (1) and (2) ensure that most of the selected galaxies are contained in the group catalog of \cite{Yang2007} (with extension to DR7). Criterion (3) guarantees that galaxies can be identified with reliable large-scale environments. Criterion (4) is to minimize the completeness problem for less massive galaxies (see \cite{Chen2019} for more details).
The reconstruction region is restricted to the redshift range of $0.01<z<0.12$, where groups with $\lg M_{halo} \gtrsim 12$ are complete. We remove the galaxies with small fraction of their virial volumes contained within the survey boundary ($0<f_{edge}\le0.6$). Our final sample contains $\sim$ 0.3 million galaxies. For these galaxies, stellar masses $M_*$, in units of $h^{-2} M_{\sun}$, are computed using the relations between the stellar mass-to-light ratio and the (g-r) color as given in \cite{Bell2003}, adopting an initial mass function (IMF) according to \cite{Kroupa2001}. We refer to \cite{Yang2007} for details.

\subsubsection{Quenched and star forming galaxies}

\label{sec_data_sfr}




\begin{figure} 
\centering
\includegraphics[width=\textwidth]{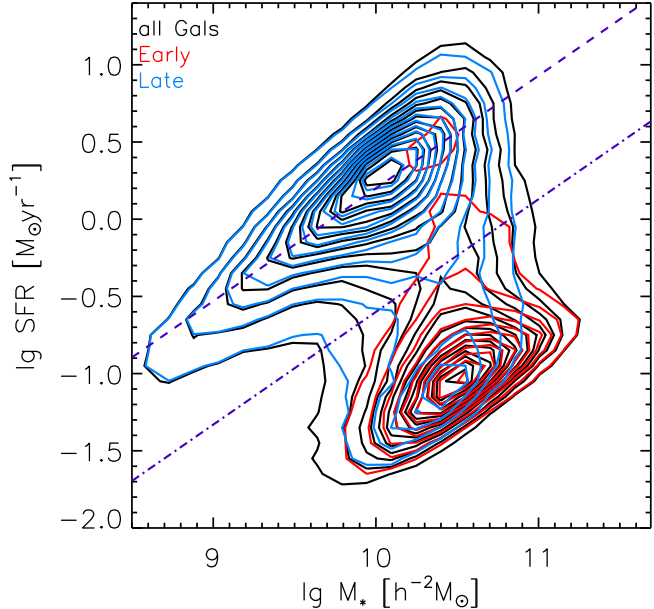}
\caption{Distributions of galaxies in the $\lg SFR - \lg M_*$ diagram, shown in contours. Early type galaxies are shown in red contours and late type galaxies are shown in blue contours. The black contours are for galaxies of all morphologies (early$+$late). The purple dashed line 
goes through the highest density regions of the MS. The purple dashed-dotted line is Equation \ref{e_sep}, which roughly pick up the lowest density regions of the contour of the full sample. We use this dashed-dotted line as the separation between the MS and the quenched populations. 
}
\label{f_CMD}
\end{figure}

We split our galaxies into quenched galaxies (qGals) and star-forming galaxies (sGals) according to their positions on the $\lg SFR - \lg M_*$ diagram (Figure \ref{f_CMD}). Here the SFRs are taken from the MPA-JHU DR7 catalog, estimated with an updated version of the method presented in \cite{Brinchmann2004} and calibrated using the Kroupa IMF \citep{Kroupa2003}. 
The bulge over total luminosity ratio B/T is taken from \cite{Simard2011}. 
Following \cite{WangE2018} and \cite{Bluck2016}, we use a line with the slope of 0.73 to separate the main sequence (MS) from the quenched populations: 
\begin{equation}
\lg SFR = 0.73 * \lg M_*- 1.46 * \lg h - 8.1.
\label{e_sep}
\end{equation}
This separation line (dashed-dotted purple line) is shifted down by 0.8 dex from the line that goes across the densest regions of the MS (dashed purple line). The separation roughly picks up the lowest density regions of the contours.

We have examined the results by adopting (1) a division with the slope of 0.73 but at slightly different positions
, (2) a flatter division from \cite{Woo2013}. Our main results hold for both (1) and (2).

We note here that the SFRs may suffer from the aperture correction problem, especially for the low redshift galaxies \citep[e.g.][]{Salim2016}. However, our paper is focusing on the quenched fraction instead of the absolute SFR values. Therefore, we care more about the strong bi-modal feature and the division line between the two populations in Figure \ref{f_CMD} instead of the exact SFR values. We checked that if we cut at lower redshifts, our main results hold (see Appendix \ref{sec:AppA} for more details). This implies that the aperture correction problem does not affect our results significantly.

Figure \ref{f_CMD} shows the $\lg SFR - \lg M_*$ diagram for the early types in red contours and for the late types in blue contours. The early types are dominated by the quenched population and the late types are dominated by sLs.


\subsubsection{Morphology Classification}
\label{sec_data_morph}

In addition to the quenched/star forming separation of galaxies, we also separate galaxies into different morphology types. \cite{HC2011} developed a machine learning algorithm that assigns the probabilities for each galaxy of different morphology types according to its color, total axis ratio, and concentration parameters. Instead of giving a certain type, each galaxies was calculated with four different probabilities $p_{Ell}$, $p_{S0}$, $p_{Sab}$, and $p_{Scd}$. 
The training set of this machine learning algorithm was taken from the visual classification of \cite{Fukugita2007}. 
We followed \cite{Meert2015} using a simple linear model to calculate the Hubble Types from the four probabilities of each galaxy (Equation \ref{e_hubble_type}). 
\begin{equation}
T = - 4.6p_{Ell} - 2.4p_{S0} + 2.5p_{Sab} + 6.1p_{Scd}
\label{e_hubble_type}
\end{equation}
The coefficients are calibrated using the visually classified galaxies of \cite{Nair2010} by an unweighted linear regression.
Then galaxies can be binned as early types and late types according to their Hubble Types.
\begin{equation}
\begin{aligned}
& Early: T \le 0.5 \\
& Late:  T > 0.5
\end{aligned}
\label{e_morph}
\end{equation}
The two visually classified sample used during this process -- the training set of the Machine Learning algorithm \citep{Fukugita2007} and the linear fit calibration catalog \citep{Nair2010}, both using the conventional Carnegie Atlas of Galaxies scheme \citep{Sandage1961,Sandage1994}. 
We note here that in some definitions, Sa spirals are considered as early type galaxies \cite[e.g.][]{Strateva2001}, but in this paper we followed \cite{Meert2015} and identify the galaxies with $-3<T
\le0.5$ as S0s, and those with $0.5<T\le4$ as Sabs. We consider ellipticals and S0s as early type galaxies, and Sab galaxies and Scd galaxies as late type galaxies.

\subsection{Environments}
\label{subsec_envir}

In this study, we probe the following environmental factors that can impact the morphological transformation and quenching of galaxies: 
on small scales: the halo mass $M_{halo}$, the halo centric radius $R_p/r_{180}$ and the third nearest neighbor distances ($d_{3nn}$); and on large scales: the cosmic web types.

\subsubsection{Small-scale Environments}
\label{sec_data_small}

The first small scale environment we specify is the halo environment that is associated with the galaxy groups. The galaxy groups used in this paper is identified by \cite{Yang2012} from the SDSS DR7, constructed with the halo-based group finder developed by \cite{Yang2005}. For each galaxy, the group catalog provides the stellar mass, the host halo mass, and the classification of being a central or a satellite galaxy. There are two method of estimating the halo masses $M_{halo}$, with the unit of $h^{-1} M_{\sun}$, in the catalog. We make use of the one that is estimated via the ranking of total stellar mass of all members brighter than $M_r = -19.5 + 5 \lg h$ in the r-band. We also assume that the most massive galaxy in a group to be the central galaxy, and all other members are referred as satellites. 

The group finder of \cite{Yang2007} is different from traditional friends-of-friends (FOF) methods in that it allows an isolated galaxy to be a group with only one member. This allows us to have the halo mass estimations for small halos that are usually located in the most underdense environments.

In addition to the halo environment, we also use the third nearest neighbor distances, $d_{3nn}$ to trace the small-scale environments.
$d_{3nn}$ is the distances to the third nearest bright galaxy ($M_r < -20.05$). $M_r=-20.05$ is the absolute magnitude limit at $z=0.12$. This cut help remove the selection bias of the preference of smaller $d_{3nn}$ values at lower z due to large number of faint galaxies observed, and the preference of larger values of $d_{3nn}$ at higher z since only more luminous galaxies can be observed further away. 


\begin{figure} 
\centering
\includegraphics[width=0.95\textwidth]{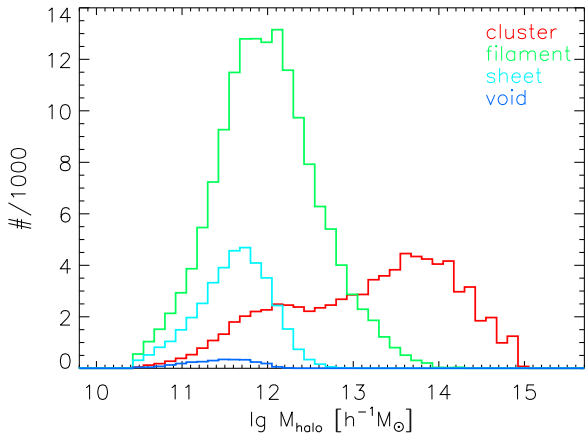}\\
\caption{Distributions of $\lg M_{halo}$ of galaxies in the four large-scale environments -- clusters (red), filaments (green), sheets (cyan), and voids (blue).
}
\label{f_lth}
\end{figure}

\subsubsection{The large-scale environments}
\label{sec_data_lss}

\begin{figure*} [!htp]
\centering
\includegraphics[width=0.45\textwidth]{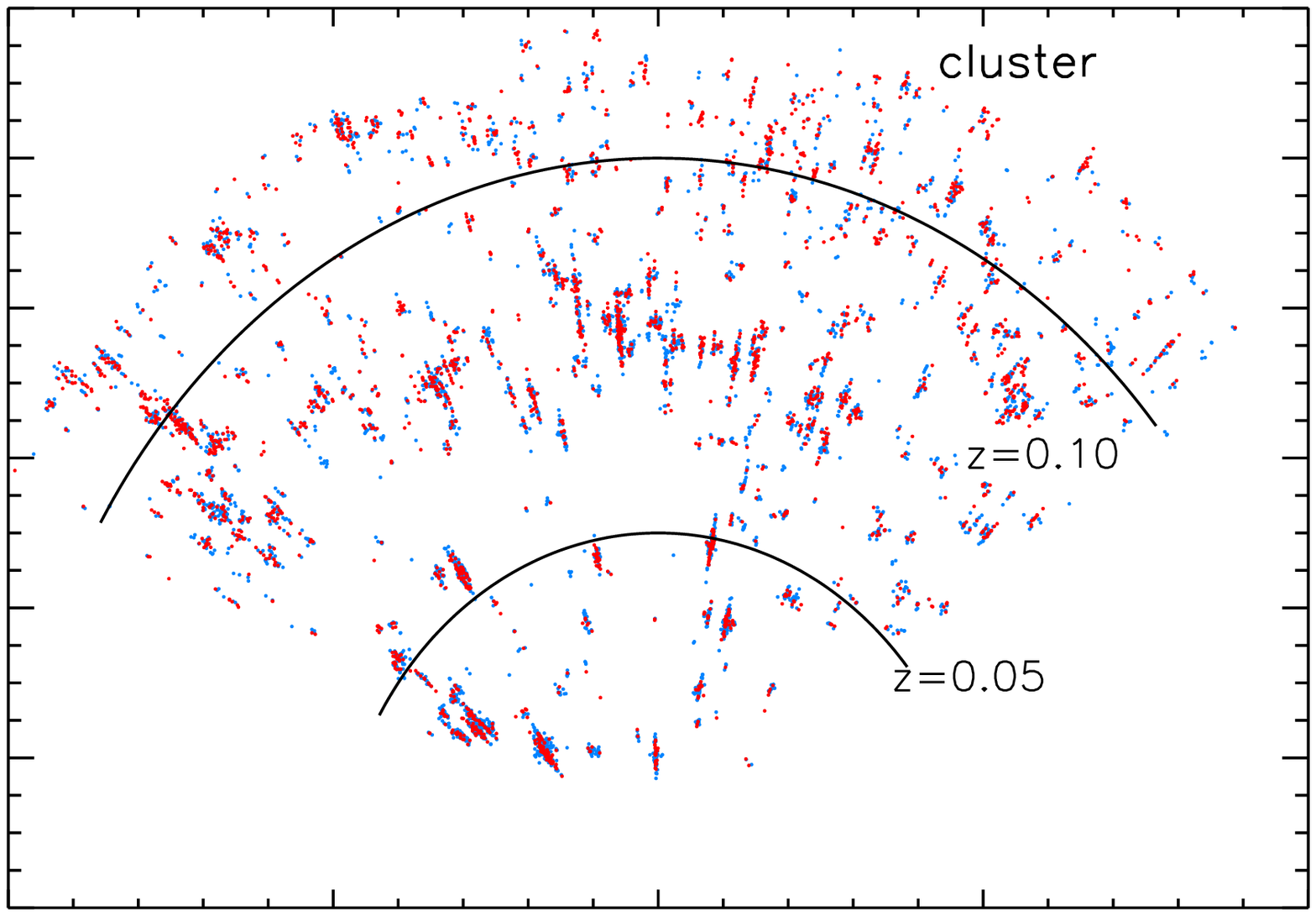}
\includegraphics[width=0.45\textwidth]{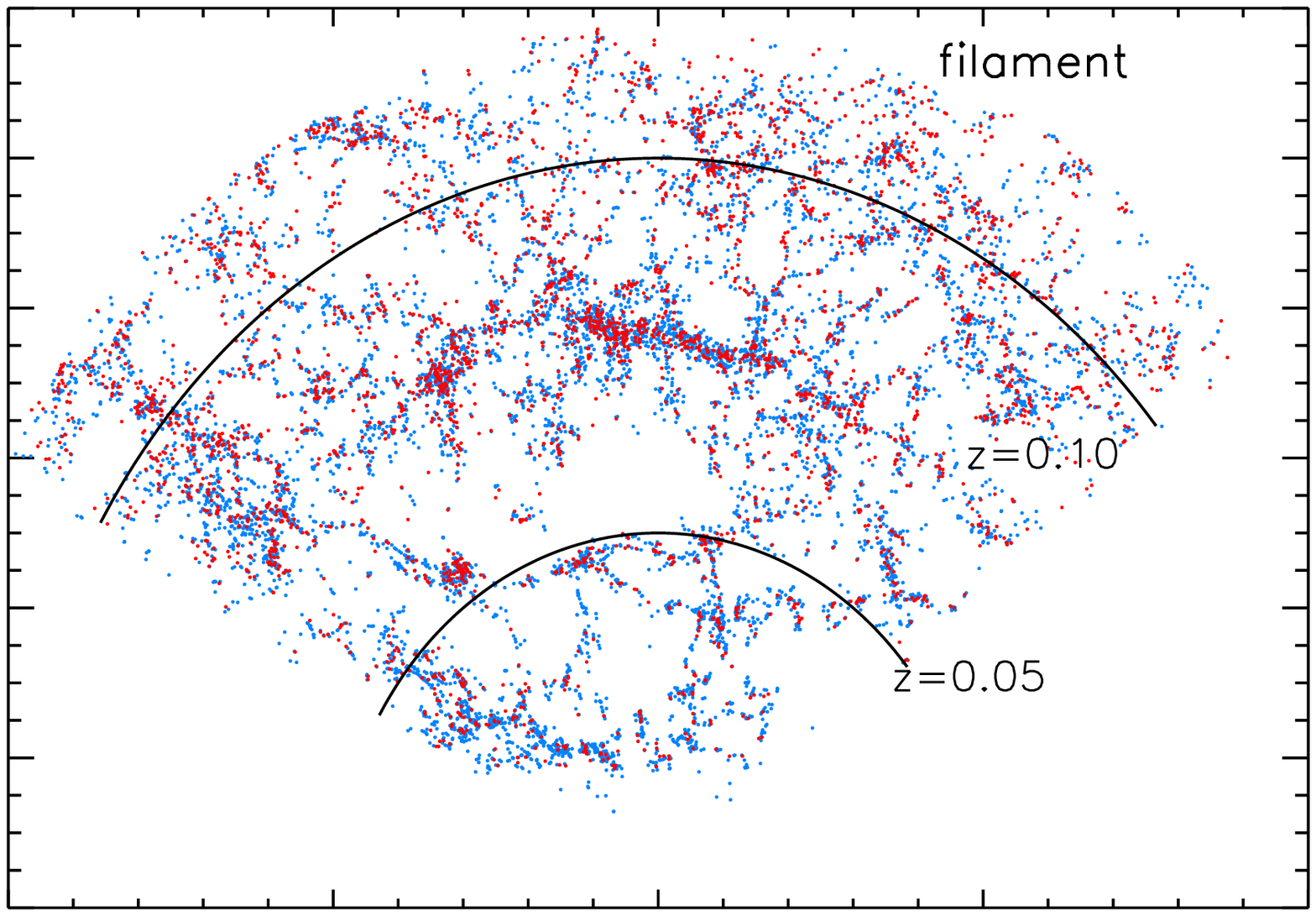}\\
\includegraphics[width=0.45\textwidth]{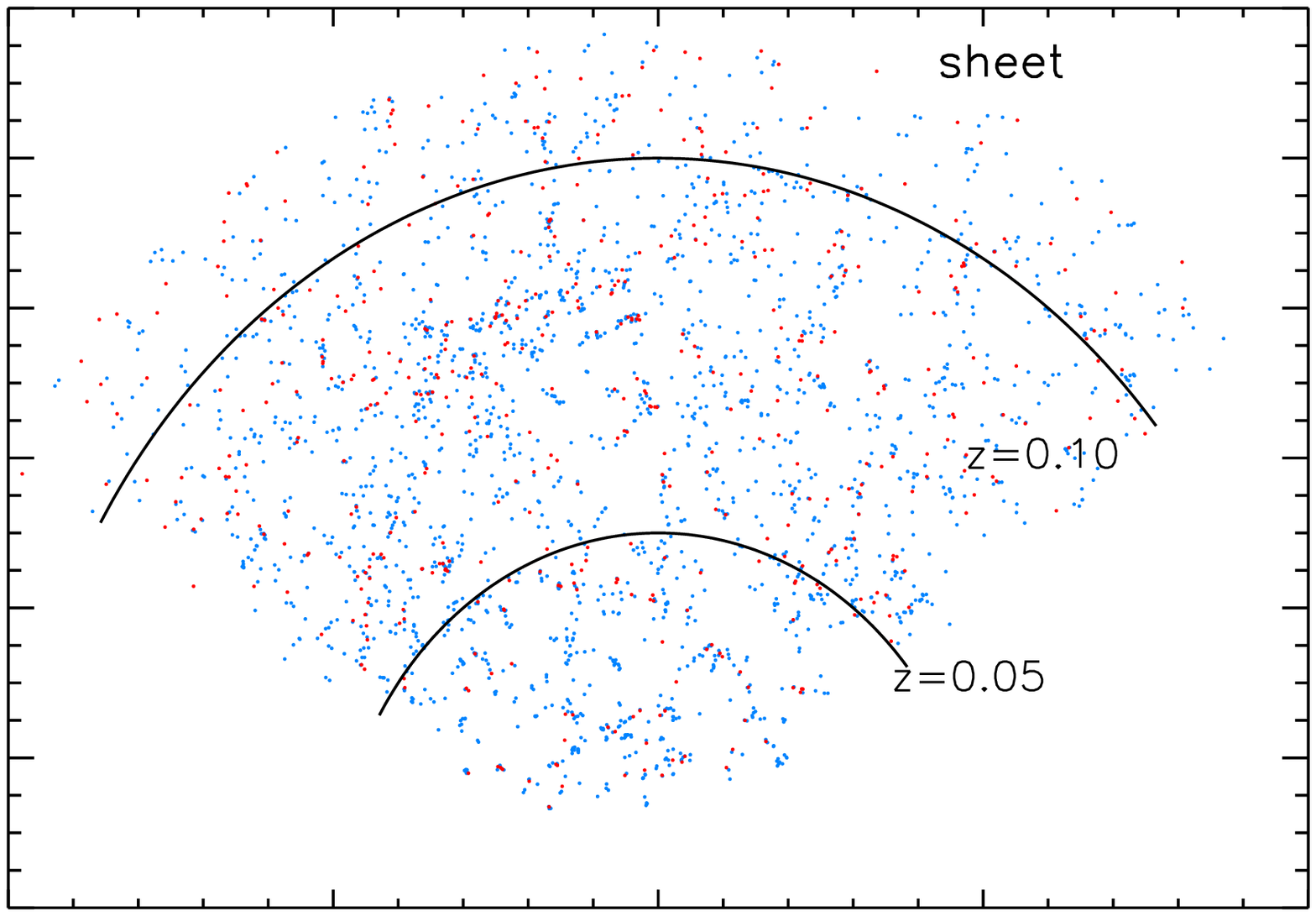}
\includegraphics[width=0.45\textwidth]{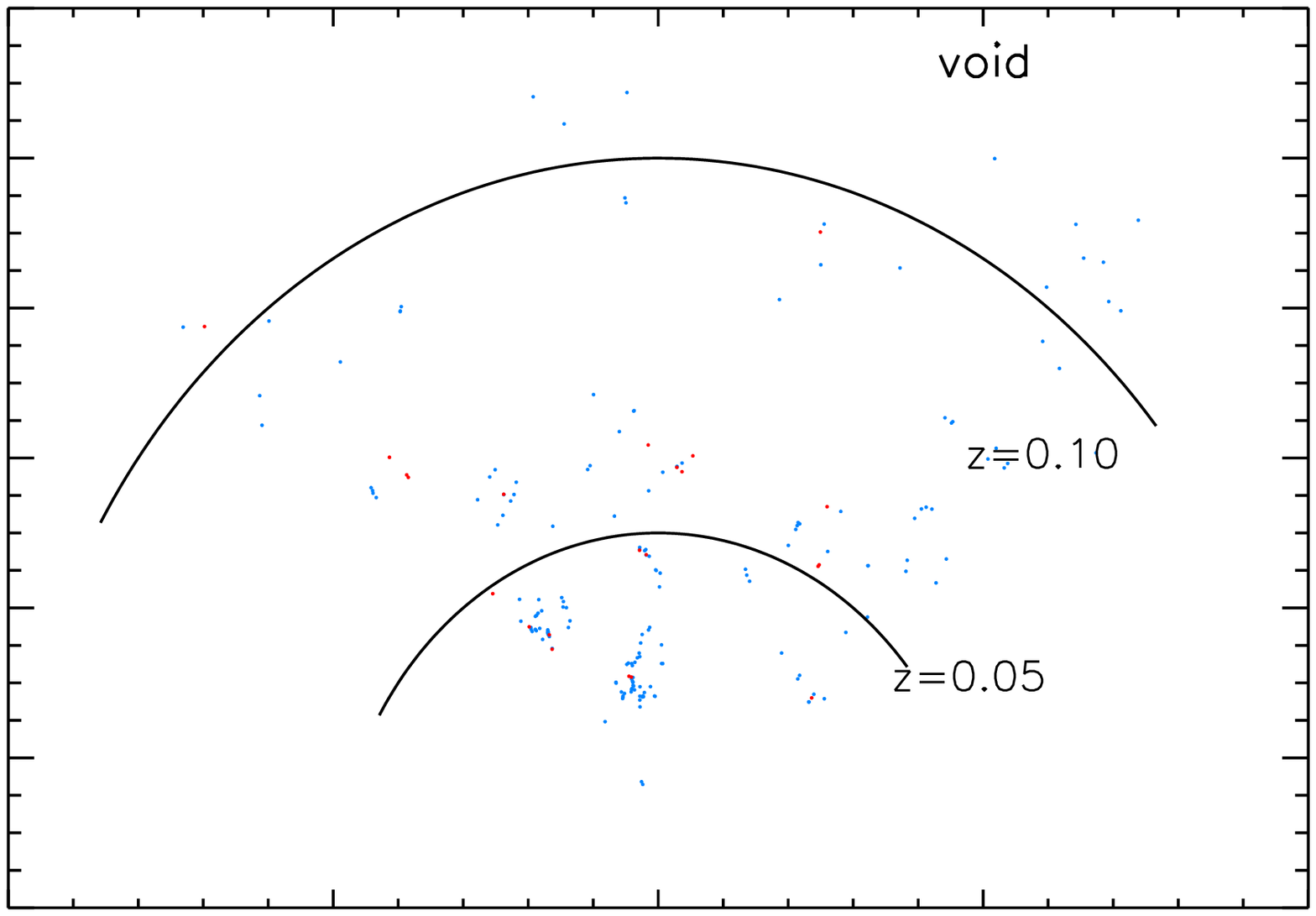}\\
\caption{Piecharts of galaxies in the four different large-scale environments in the $0\degree<dec<5\degree$ slice. Radius direction is controlled by redshifts. Orientation is controlled by RA. Only the region with $120\degree<RA<240\degree$ is shown here. Black solid circles show the positions of $z=0.05$, and $z=0.1$. Red dots are early type galaxies and blue dots are late type galaxies.. 
}
\label{f_pie_lss}
\end{figure*}

We use the cosmic web types identified by \cite{Wang2016} as large scale environments. Each galaxy was calculated with three eigenvalues ($l_1$, $l_2$, and $l_3$) following \cite{Hahn2007}. Each eigenvalue represents the tidal field tensor of each space dimension. Positive eigenvalue ($l_i>l_{th}$ and $l_{th}=0$) means that the associated space dimension is being squeezed. Therefore, the dynamical large-scale environments can be defined with the number of positive eigenvalues: If all three eigenvalues of a galaxy are positive, then it is in a cluster; Galaxies in filaments have two positive eigenvalues; One positive value defines galaxies in sheets; All three eigenvalues are negative for void galaxies. Here, $l_{th} = 0$ is set to be the threshold of the tidal tensors. We also tried other thresholds ($l_{th} = 0.2, 0.4$). The main results throughout this paper hold.

Figure \ref{f_lth} shows the distributions of $\lg M_{halo}$ for galaxies in the four different large-scale environments. Void galaxies populate the lowest range of halo masses. Galaxies in the sheets have slightly larger halo masses than those in voids, and less massive halo masses compared to those in filaments. Galaxies in clusters populate the widest range of halo masses. The most massive halos ($M_{halo}\gtrsim 10^{14}\ M_{\sun}$) are exclusively in clusters. 

Figure \ref{f_pie_lss} shows the piecharts of galaxies in clusters, in filaments, in sheets, and in voids of the declination slice of $0\degree<dec<5\degree$ in the region of $120\degree<RA<240\degree$. Galaxies are color-coded by their Hubble types calculated according to Equation \ref{e_hubble_type}. In the top left panel of Figure \ref{f_pie_lss}, galaxies in clusters are mostly distributed along the redshift direction forming short lines in the radius direction. This is the ``Finger of the God effect''. Galaxies of both early types and late types can be found in all environments.



\section{Results}
\label{sec_results}

In this section, we study the environmental effects on quenching and morphologies on different scales. To this end, we calculate the fraction of subsample $s$ out of the sample $S$ ($s\in S$) with $f_{s}(S) = \sum_{i\in \{s\}} w_{i}/\sum_{j\in \{S\}} w_{j}*100\%$.
To correct the Malmquist bias and redshift (spectroscopic) incompleteness of galaxies, we assign each galaxy with a weight $w= 1/(V_{max}C)$. The redshift completeness $C$ is taken from the NYU-VAGC. $V_{max}$ is calculated as the volume of the reconstruction region that is between $z_{min}$ and $z_{max}$, where $z_{min}$ and $z_{max}$ are the minimum and maximum redshifts, between which the galaxy can be observed with the r-band limit of 17.72 mag. We refer the readers to \cite{WangH2018,WangE2018} for more details. 
\cite{vdBosch2008} carefully compared the $v_{max}$ method with volume-limited method, and found that both methods give similar results. \cite{Chen2019} further confirmed that the $v_{max}$ method is appropriate to be applied to galaxy samples with $\log M_* \ge 9$.

\subsection{The large-scale environments}
\label{sec_result_lss}

\begin{table} 
\centering
\caption{Abundances of early and late type galaxies\label{t_lss}}
\begin{tabular}{c|cc|cccc|cc}\hline\hline
f(\%) & early & late & sE & qE & sL & qL & quench & MS\\\hline
all&15.7&84.3& 4.1&11.6&67.2&17.0&28.6&71.4\\\hline
cluster&23.7&76.3& 3.5&20.1&46.8&29.5&49.6&50.4\\\hline
filament&14.2&85.8& 4.3& 9.9&72.1&13.7&23.6&76.4\\\hline
sheet& 9.7&90.3& 4.4& 5.3&80.6& 9.7&15.0&85.0\\\hline
void& 8.2&91.8& 5.1& 3.1&83.7& 8.1&11.2&88.8\\\hline\hline
\end{tabular}\\
\end{table}

\begin{figure} 
\centering
\includegraphics[width=0.95\textwidth]{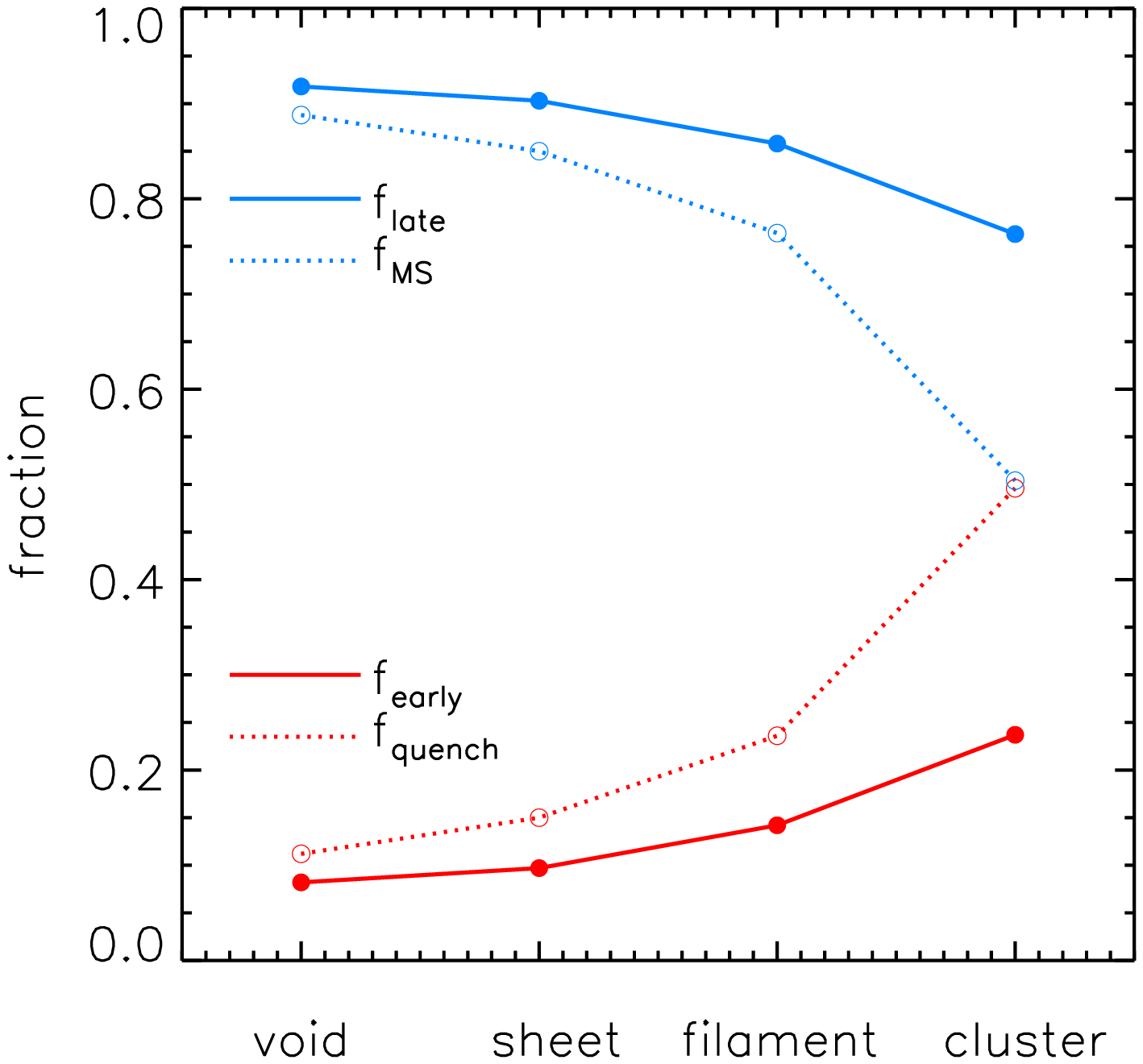}\\
\caption{The fraction of early type galaxies out of all morphological types of galaxies ($f_{early}$, shown in red solid line) and the late type fraction out all morphological types of galaxies ($f_{late}$, shown in blue solid line) in the four large-scale structures. The fraction of the quenched galaxies out of galaxies of any type of star-forming state ($f_{quench}$, shown in red dotted line) and the fraction of the MS out of galaxies of any type of star-forming state ($f_{MS}$, shown in blue dotted line) in the four large-scale environments.
}
\label{f_tab1}
\end{figure}

We first calculate the fraction of subsample galaxies out of the total population for all galaxies, in different cosmic web environments. Table \ref{t_lss} shows the fractions of the four types of galaxies out of all galaxies in the four large-scale structures. We also show these fractions in Figure \ref{f_tab1}. For early types, the fraction of qEs out of all galaxies in all large-scale structures is 11.6\%, much higher than the fraction of sEs out of all galaxies(4.1\%). The statistics of early types is therefore dominated by qEs, which results in the notion that early types are always thought to be red and passive. The overall fractions of sLs and qLs out of all galaxies in all large-scale environments are 67.2\% and 17.0\%. Late types are dominated by the star-forming subsample. Thus, they are usually thought to be actively star-forming. 

The morphology-density relation can be seen. The fractions of early types out of all cluster galaxies decreases from 23.7\% all the way through filaments and sheets, and get to 8.2\%, which is the fraction of early types out of all void galaxies; the fraction of late type galaxies out of all cluster galaxies gradually increases from 76.3\% to 91.8\%, which is the fraction of late type galaxies out of all void galaxies.
Table \ref{t_lss} also shows the SFR-density relation. The fraction of quenched galaxies out of all void galaxies increases from 11.2\% to 49.6\%, which is the fraction of quenched galaxies out of all galaxies in clusters.
It's interesting to note that the star formation properties change much more dramatically than morphology as environment changes (Figure \ref{f_tab1}).

In early types, the sE fraction ($f_{sE}$) is almost constant in all four large-scale structures ($\sim 4\%$).
However, the qE fraction $f_{qE}$ decreases strongly from clusters to voids. In late types,  $f_{sL}$ increase while $f_{qL}$ decrease significantly from clusters to voids.

With all these statistics with respect to the large scale environments, we will discuss the environmental effects on the morphological transformation and the quenching with more details in Section \ref{sec_result_small} and Section \ref{sec_discuss}.

\subsection{The in-situ and small-scale environments}
\label{sec_result_small}

There is a degeneracy between the large-scale environments and the in-situ and small-scale environments. Galaxies in clusters, the densest large-scale environments, tend to have more massive central galaxies, be hosted by the most massive halos (Figure \ref{f_lth}) and have the smallest separations with their neighbors (smallest $d_{3nn}$ values). Galaxies in voids, the most under-dense large-scale environments, tend to have low mass central galaxies, and the least massive halos (Figure \ref{f_lth}) and farthest away from their neighbors (largest $d_{3nn}$ values). In order to disentangle the degeneracy between the large- and small- scale environments, it is therefore reasonable to study the large-scale environmental effects with the small-scale environmental parameters controlled. 

\begin{figure} 
\centering
\includegraphics[width=\textwidth]{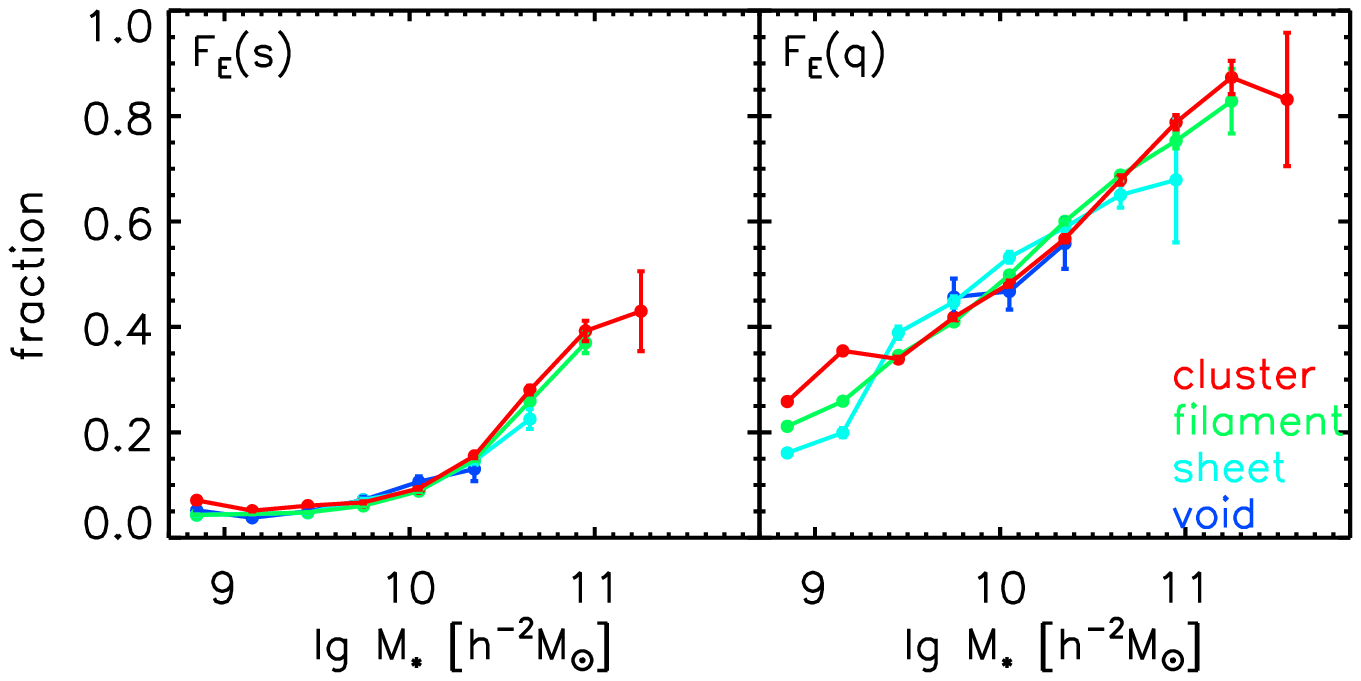}\\
\includegraphics[width=\textwidth]{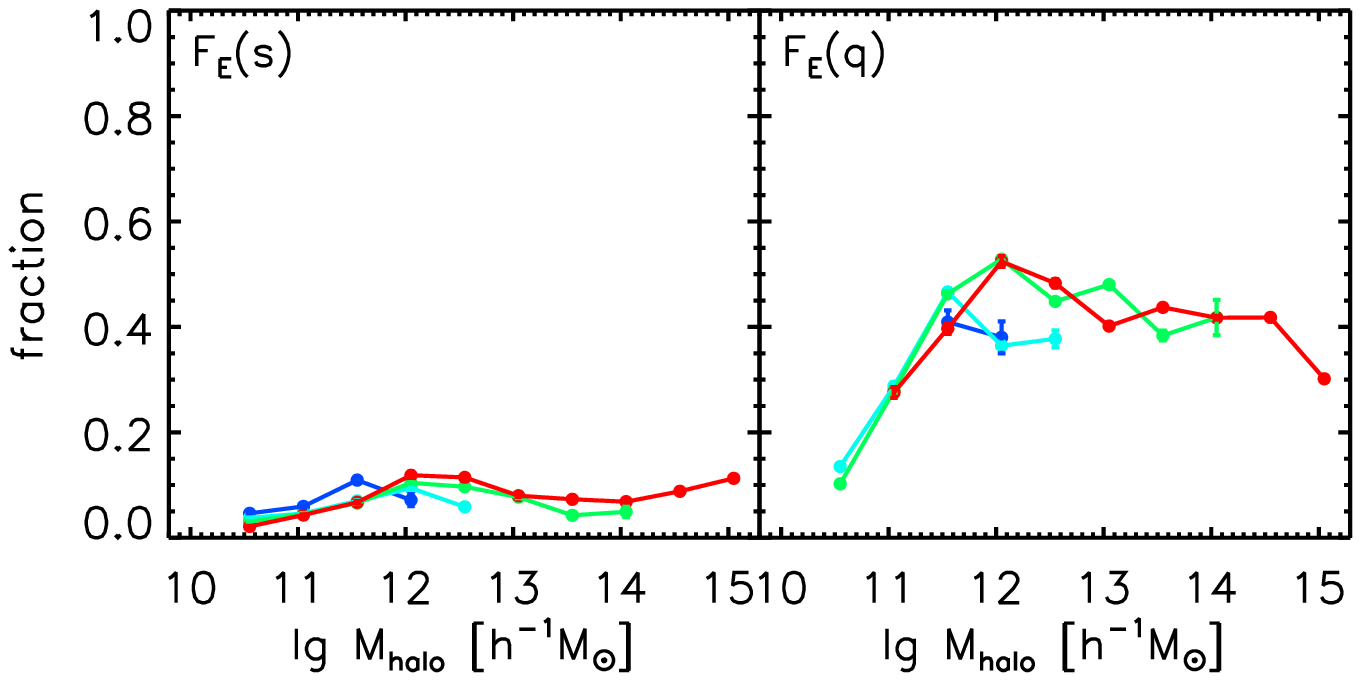}\\
\includegraphics[width=\textwidth]{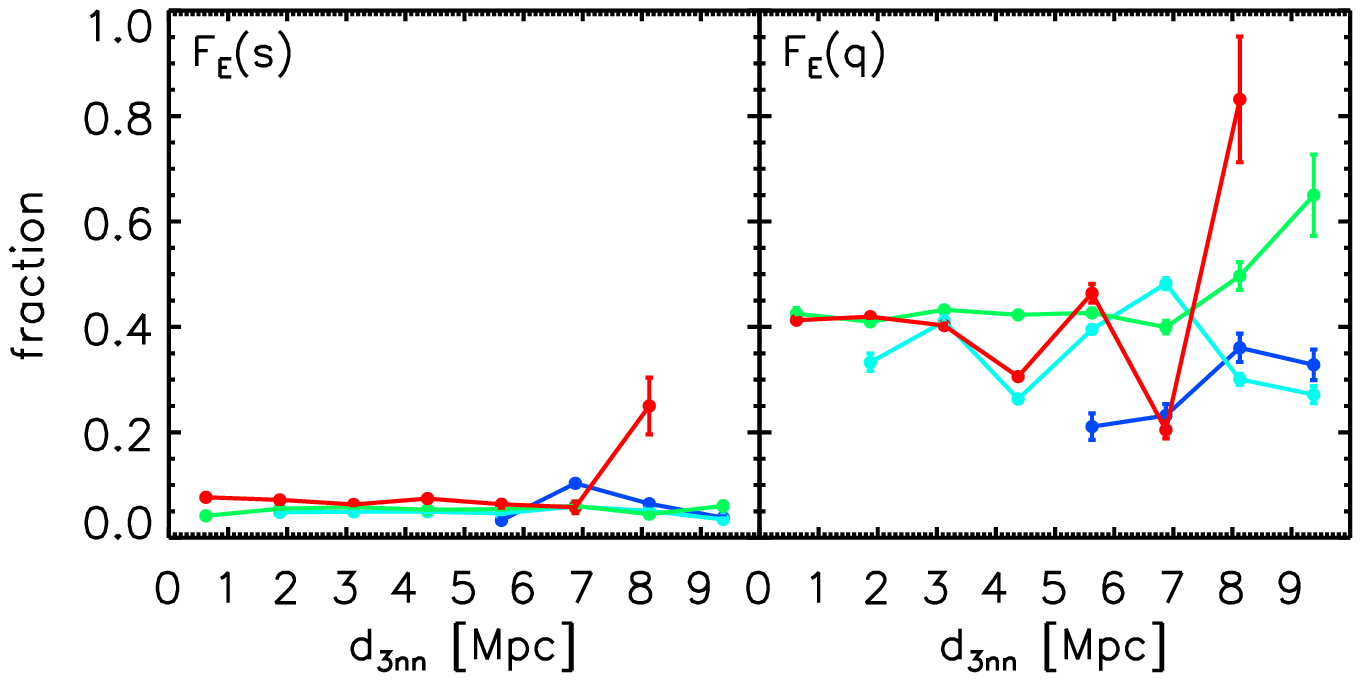}\\
\caption{The fraction of early type star-forming galaxies out of all star-forming galaxies (left, $F_E(s)$) and the fraction of the early type quenched galaxies out of all quenched galaxies }(right, $F_E(q)$) as a function of the small-scale parameters. Top to bottom: the stellar mass $\lg M_*$, the halo mass $\lg M_{halo}$, and the third nearest neighbor distances $d_{3nn}$. Red is for galaxies in clusters, green is for galaxies in filaments, cyan is for galaxies in sheets, and blue is for galaxies in voids. Only bins with $n\ge50$ are shown. Errors are estimated using Poisson errors ($\sqrt{n}$). Error bar for $f=n_a/n_b$ is then $\sigma(f)=\sqrt{n_a/n_b^2+n_a^2/n_b^3}$.

\label{f_small_FE}
\end{figure}

\begin{figure} 
\centering
\includegraphics[width=\textwidth]{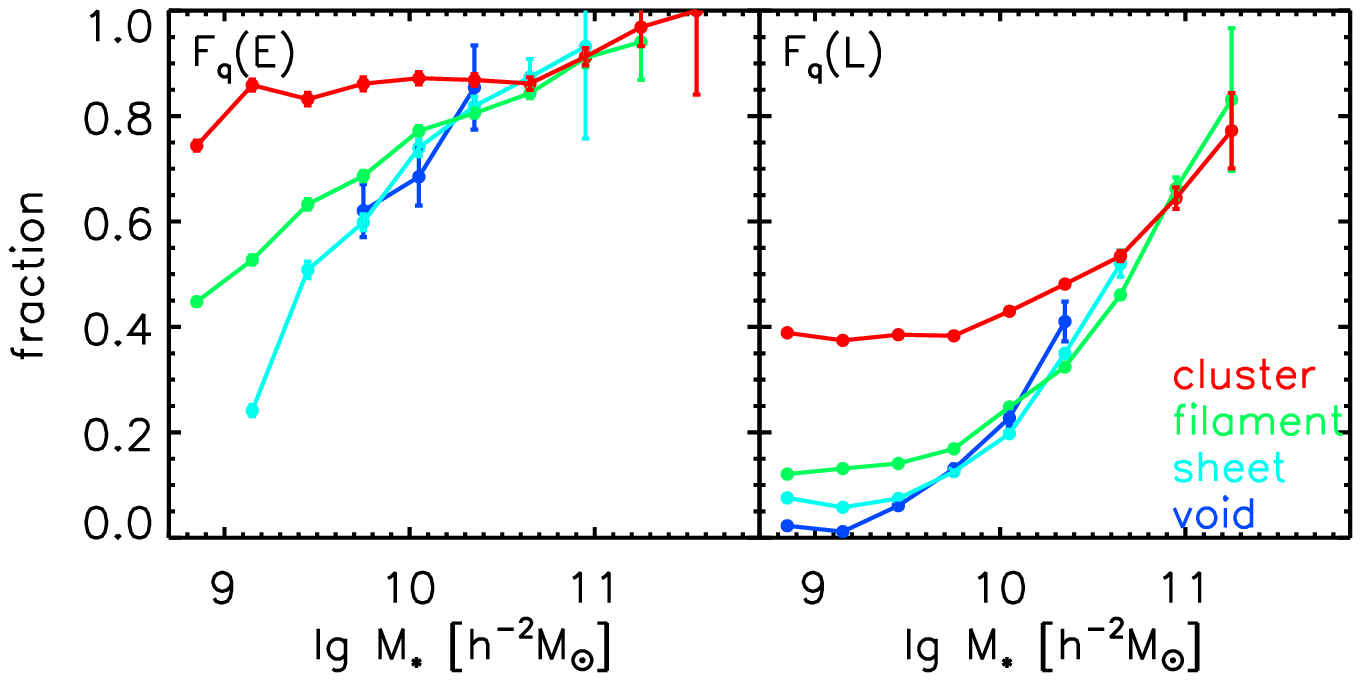}\\
\includegraphics[width=\textwidth]{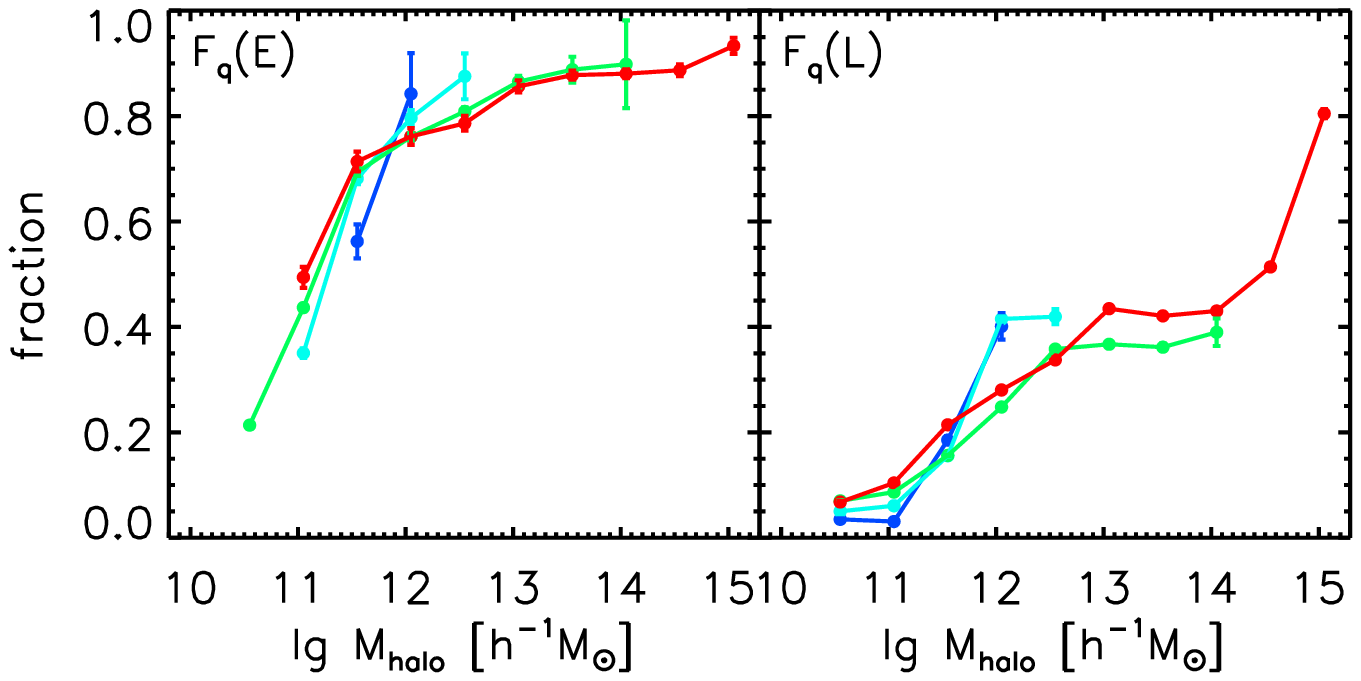}\\
\includegraphics[width=\textwidth]{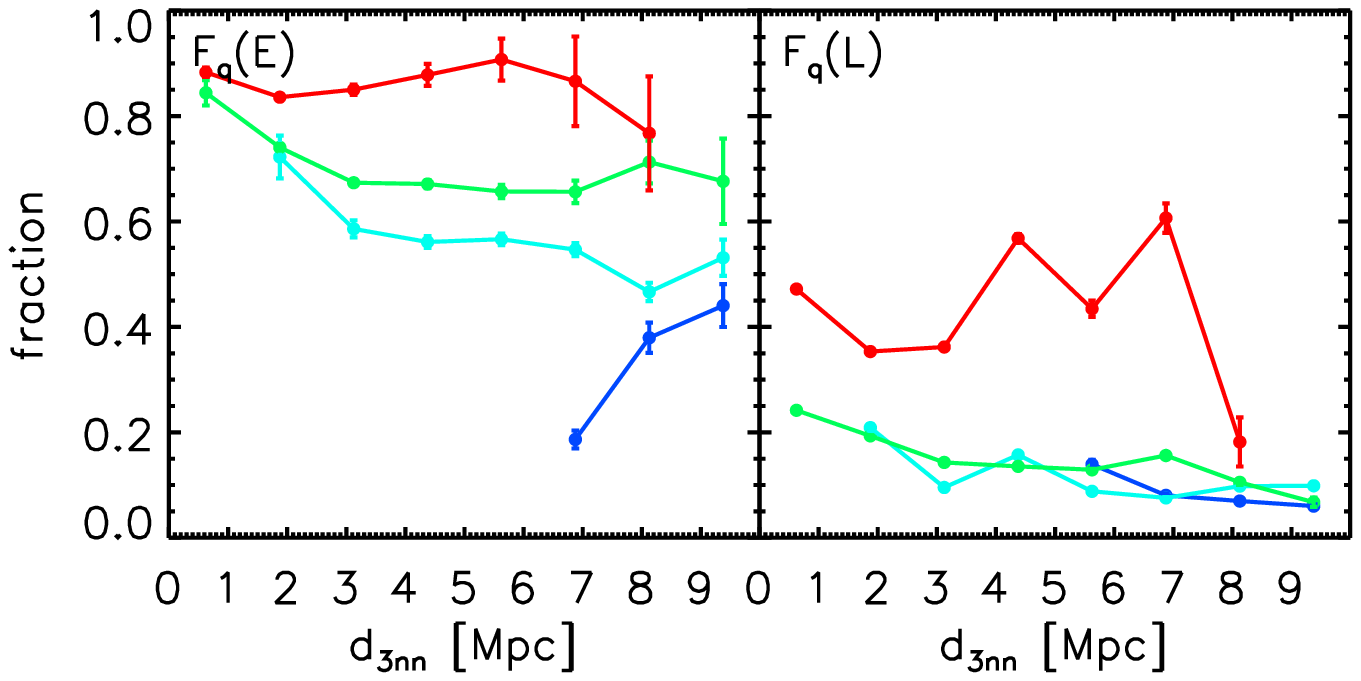}\\
\caption{Similar plots with Figure \ref{f_small_FE}, but for the quenched fraction $F_q$ out of all early type galaxies (left, $F_q(E)$) and the quenched fraction out of all late type galaxies (right, $F_q(L)$).
}
\label{f_small_Fq}
\end{figure}

We study the in-situ and small-scale environments with three different parameters ($M_*$, $M_{halo}$ and $d_{3nn}$) in Figure \ref{f_small_FE} and Figure \ref{f_small_Fq}. Each of these parameters represents an in-situ or small-scale environment at different scale: $M_*$ represents the internal physical processes on the galaxy-scale, $M_{halo}$ is a gravity bounded halo-scale parameter; $d_{3nn}$ is the distance to the third nearest bright neighbor (see Section \ref{sec_data_small} for more details of the definition). 

\begin{itemize}
    \item morphological transformation
\end{itemize}

We first study the morphological transformation of galaxies in different cosmic web environments. The resulting early type galaxy fractions are shown in Fig. \ref{f_small_FE}. Among the star-forming MS galaxies (the left panels), the fraction of early type galaxies increase as a function of $M_*$, an in-situ environmental factor. The fraction does not show strong dependence on small-scale environments, $M_{halo}$ and $d_{3nn}$. In addition, the galaxies in different cosmic web environments show quite consistent behaviors. That is, for the star-forming main sequence galaxies, the morphological transformation, which only depends on $M_*$, is an in-situ process.

For the quenched galaxies (the right panels), quite similar to the star-forming galaxies, there is a strong dependence of early type fraction on the in-situ parameter, $M_*$. There are no small-scale and large-scale environmental dependencies, again indicating that the  morphological transformation is an in-situ process. On the other hand, we also notice that, with the same stellar mass $M_*$, the quenched galaxies have overall higher early type fractions than those of star-forming galaxies. This difference indicates that the quenching and morphological transformation are correlated. 

\cite{Bamford2009} studied the early-type fraction as a function of local density for galaxies in various stellar mass bins in the left panel of their Figure 12. They found that the stellar mass is very important for the early type fractions, with the highest mass bin of early type fraction ($>$60\%) three times higher than that of the lowest mass bin ($<$20\%). This is consistent with what we found in the top panels of Figure \ref{f_small_FE}. We will discuss more about the comparison between our results and theirs in Section \ref{sec_dis_relations}.

\begin{itemize}
    \item quenching
\end{itemize}

Next, we show the quenched fraction of galaxies in different cosmic web environments in Figure \ref{f_small_Fq}. Among the early type galaxies (left panels), the quenched fraction varies as a function of $M_*$ and $M_{halo}$. While the dependence on $d_{3nn}$ is not very significant. More importantly, we see that the $M_{halo}$ dependence of the quenched fractions is quite consistent for galaxies in different cosmic web environments.  
Although we see that the quenched fractions as a function of $M_*$ and $d_{3nn}$ are very different for galaxies in different cosmic web environments, these large scale environmental dependencies could be caused by the correlation between the halo mass and cosmic web types. Shown in the right panels of Figure \ref{f_small_Fq} are results for the late type galaxies. The  general trends, i.e., the $M_*$, $M_{halo}$ and  $d_{3nn}$  dependencies are quite similar to those of early type galaxies. Quantitatively, we see that the late type galaxies have overall lower quenched fractions, again indicating that the quenching and morphological transformation are dependant. These behaviors suggest that the quenching of galaxies is possibly only impacted by in-situ and small-scale halo environments, not large-scale environments, in very good agreement with those findings obtained in \cite{WangH2018}.

Let's get back to the first panel of Figure \ref{f_small_Fq}. If we attribute the large-scale environmental dependence to the halo mass dependence, then the significant differences seen in the quenched fraction of low mass galaxies in different cosmic web environments suggests that for low mass galaxies,  quenching is mainly driven by halo environments. The higher quenching fraction in denser cosmic web structures for less massive galaxies is the result of the high quenching ratio of satellites in more massive halos ($M_{halo}\ga 10^{13.0}\msunh$). We will re-visit this problem in Section \ref{sec_dis_quench} with Figure \ref{f_Fmorph}.
While for massive galaxies, the quenching fraction is consistent in all four cosmic web structures, which suggests to be stellar mass quenching dominated.

\section{Results on the $\lg M_{halo}-\lg M_*$ map}
\label{sec_discuss}

In Section \ref{sec_result_small}, we show that both the morphological transformation and the quenching do not vary significantly with $d_{3nn}$ (the bottom panels of Figure \ref{f_small_FE} and Figure \ref{f_small_Fq}). When the stellar mass is controlled, the early type fraction is consistent in all four cosmic web structures (the top panel of Figure \ref{f_small_FE}). When the stellar mass is controlled at high stellar masses or when the halo mass is controlled at low halo masses, the quenching fraction is consistent in all large-scale environments (the top panel and the middle panel of Figure \ref{f_small_Fq}).
This strongly suggests that the nearest neighborhoods and the large-scale environments are neither important in the morphological transformation from late types to early types, nor important in the quenching of SF. \textbf{The morphological transformation is mainly regulated by the stellar mass. Quenching is mainly driven by the stellar mass for more massive galaxies and by the halo mass for galaxies with smaller stellar masses.}
As SF quenching and morphological transformation are correlated, in this section, we further explore these two processes and their relation with stellar masses and halo masses on 2-D maps, especially, the $\lg M_{halo}-\lg M_*$ map. 

\subsection{The morphological transformation}
\label{sec_dis_indepen}

\begin{figure*}[!htp]
\centering
\includegraphics[width=0.9\textwidth]{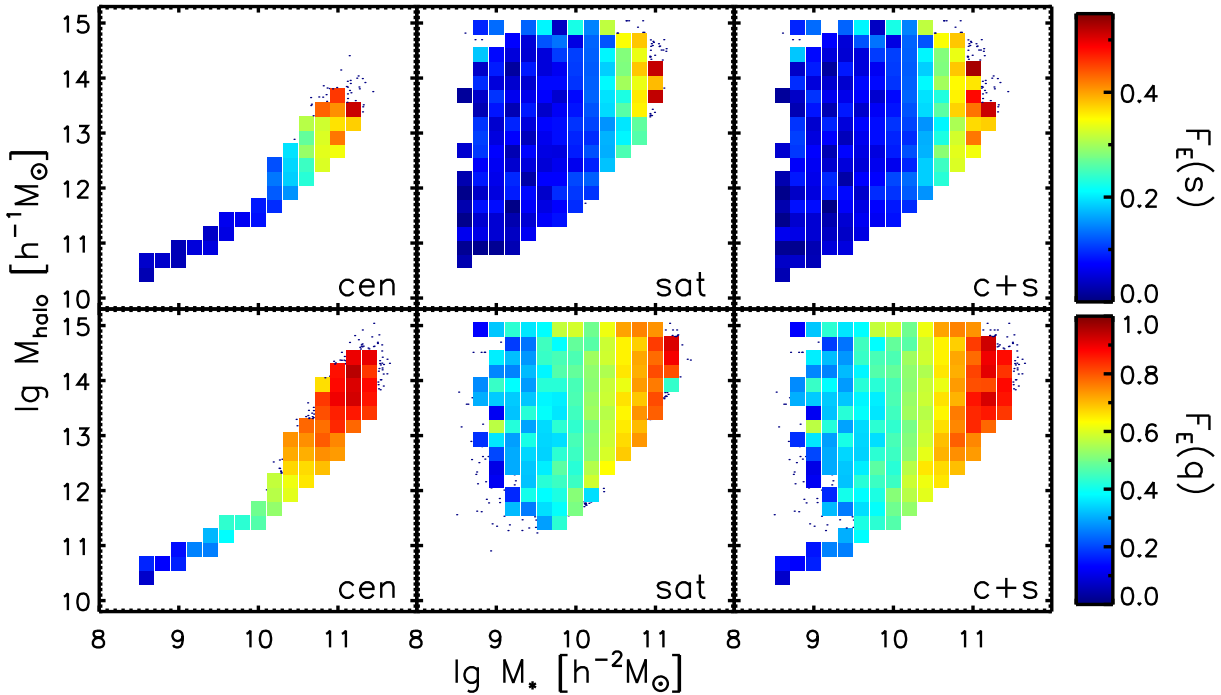}\\
\caption{Early type fractions at fixed SF state on the $\lg M_{halo}-\lg M_*$ diagram. Bins are color coded by their early type fractions, as are shown by the color bars. Upper panels are for the fraction of early type galaxies out of all MS galaxies ($F_E(s)$) and lower panels are for the fraction of early type galaxies out of all quenched galaxies ($F_E(q)$). Left to right: centrals, satellites, and (cen$+$sat). Only bins with $n(MS)\ge20$ for the upper panel and $n(quenched)\ge20$ are shown for the bottom panel. For the other bins that do not satisfy the above criteria, we show the positions of the early types as black dots.
}
\label{f_Fqs}
\end{figure*}

We first study the morphological transformation from late types to early types on the $\lg M_{halo}-\lg M_*$ map in Figure \ref{f_Fqs} for star forming main sequence and quenched galaxies separately. Bins in Figure \ref{f_Fqs} are color coded by the early type fraction ($f_E$). The upper panel is for the fraction of the early type galaxies out of all MS galaxies ($f_E(s)$), and the lower panel is for the fraction of early type galaxies out of all quenched galaxies ($f_E(q)$). We also split the centrals and the satellites in the first and second columns, and the third column is the (cen$+$sat) sample. 

We find that for both MS and the quenched population, bins with similar early type fraction (similar color coding) are roughly vertically distributed, especially for the satellites (middle column). For the centrals (left column), since the stellar masses and halo masses are in tight correlation separately, it is hard to disentangle their contribution on the $\lg M_{halo}-\lg M_*$ map. This suggests that the morphological transformation is mostly regulated by stellar masses and almost independent from the halo masses. 
In addition, as we have separate galaxies into sub-samples of the star-forming main sequence and the quenched galaxies, we see that the quenched galaxies have overall much higher fractions of early type galaxies.

\begin{figure*}[!htp]
\centering
\includegraphics[width=0.45\textwidth]{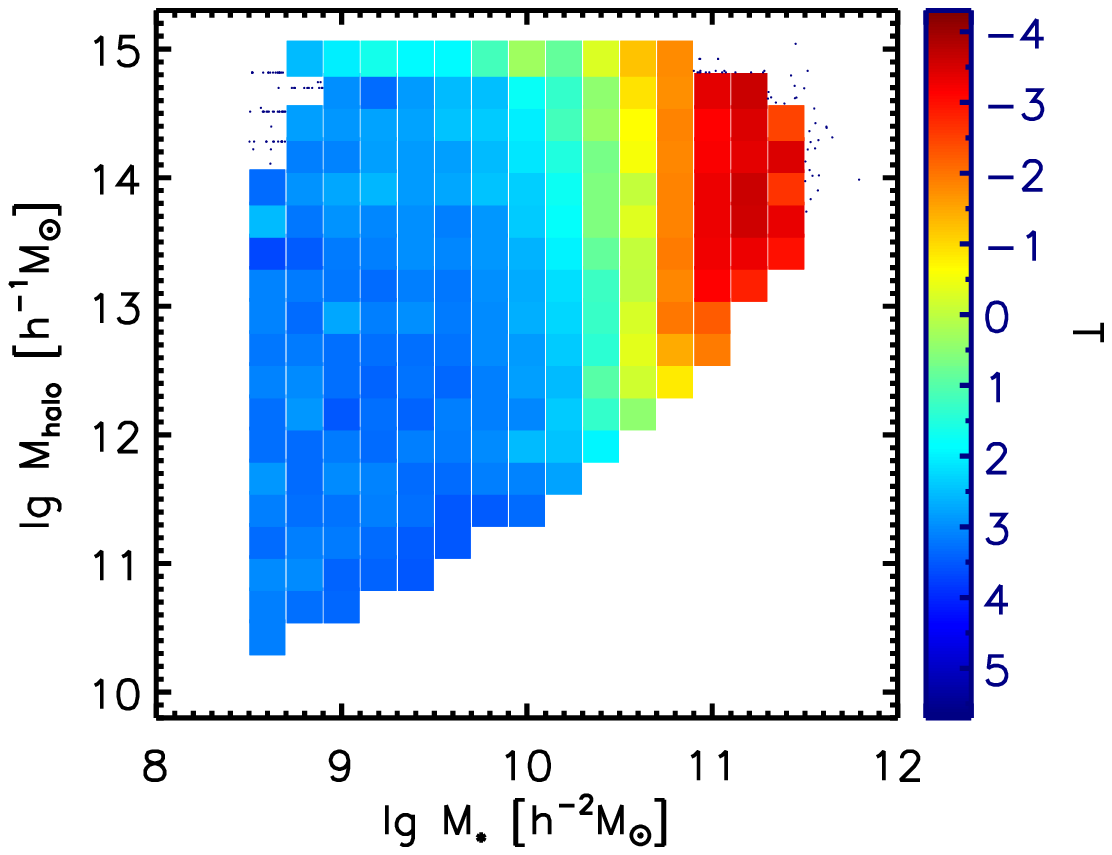}
\includegraphics[width=0.45\textwidth]{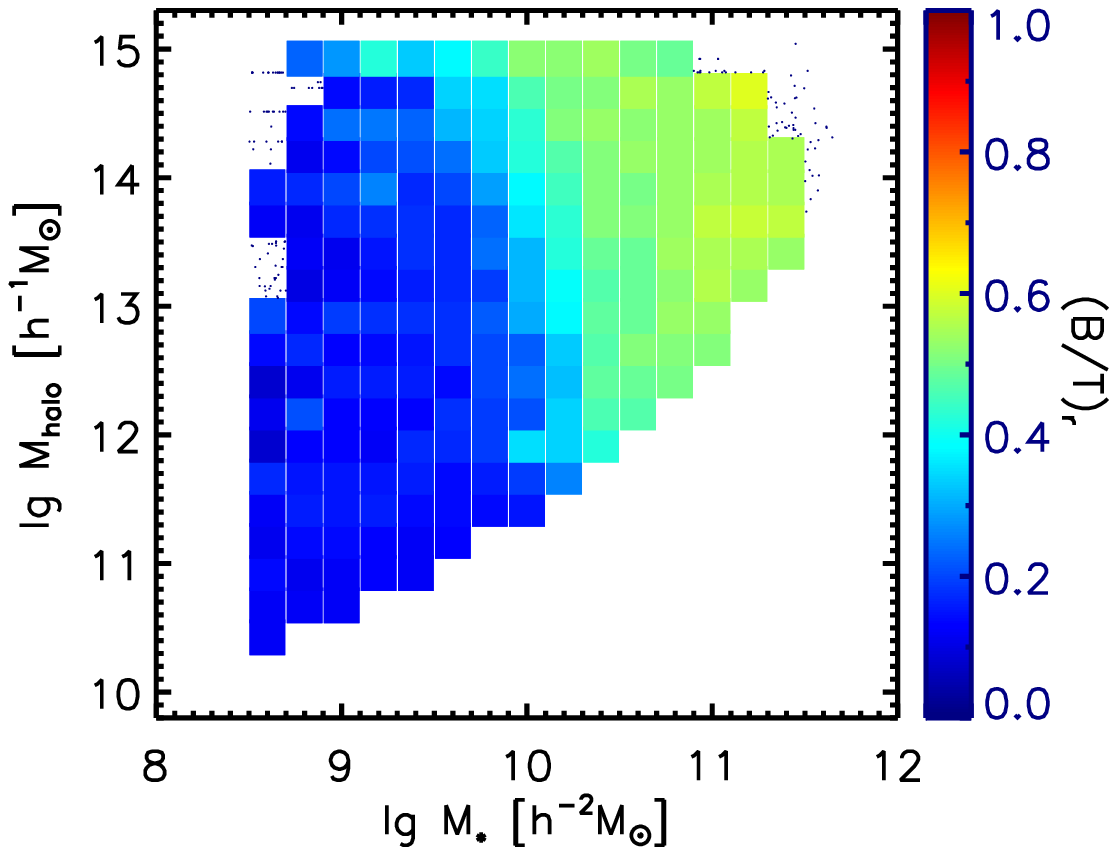}\\
\caption{The $\lg M_{halo} - \lg M_*$ maps color coded by the median Hubble T types (left) and the median B/T of each bin. Only bins with $n(gal)\ge20$ are shown. For bins with $n(gal)<20$, galaxies are shown as black dots.
}
\label{f_Mhs}
\end{figure*}

In addition to the early type fractions, we probe another two parameters that are associated with the morphological transformation of galaxies, i.e., the Hubble types and the bulge-to-total light ratios. Figure \ref{f_Mhs} shows the distributions of all galaxies on the $\lg M_{halo} - \lg M_*$ map. In the left panel, bins are color coded by their median Hubble types T calculated by Equation \ref{e_hubble_type}. In the right panel, bins are colored by their median bulge-to-total light ratio, (B/T)s. These two panels together give the information of how significantly the stellar mass and how negligibly halo mass affect the morphological transformation. We note that although the changing of color in the bulge-to-total light ratio panel is not that purely vertical as the Hubble types T,  in both panels, the color changes mainly in the stellar mass direction. This confirms our result in Section \ref{sec_result_small} that the morphological transformation is mainly regulated by the stellar mass. 

\begin{figure*}[!htp]
\centering
\includegraphics[width=0.45\textwidth]{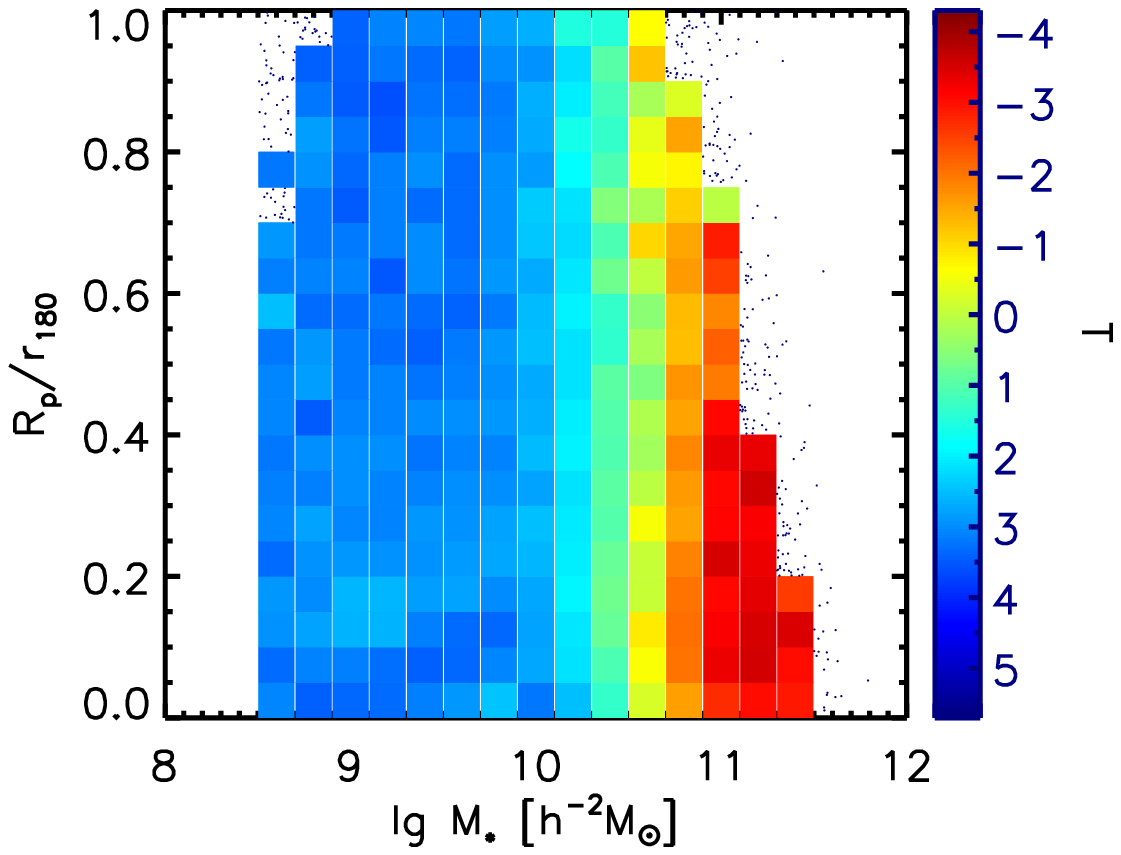}
\includegraphics[width=0.45\textwidth]{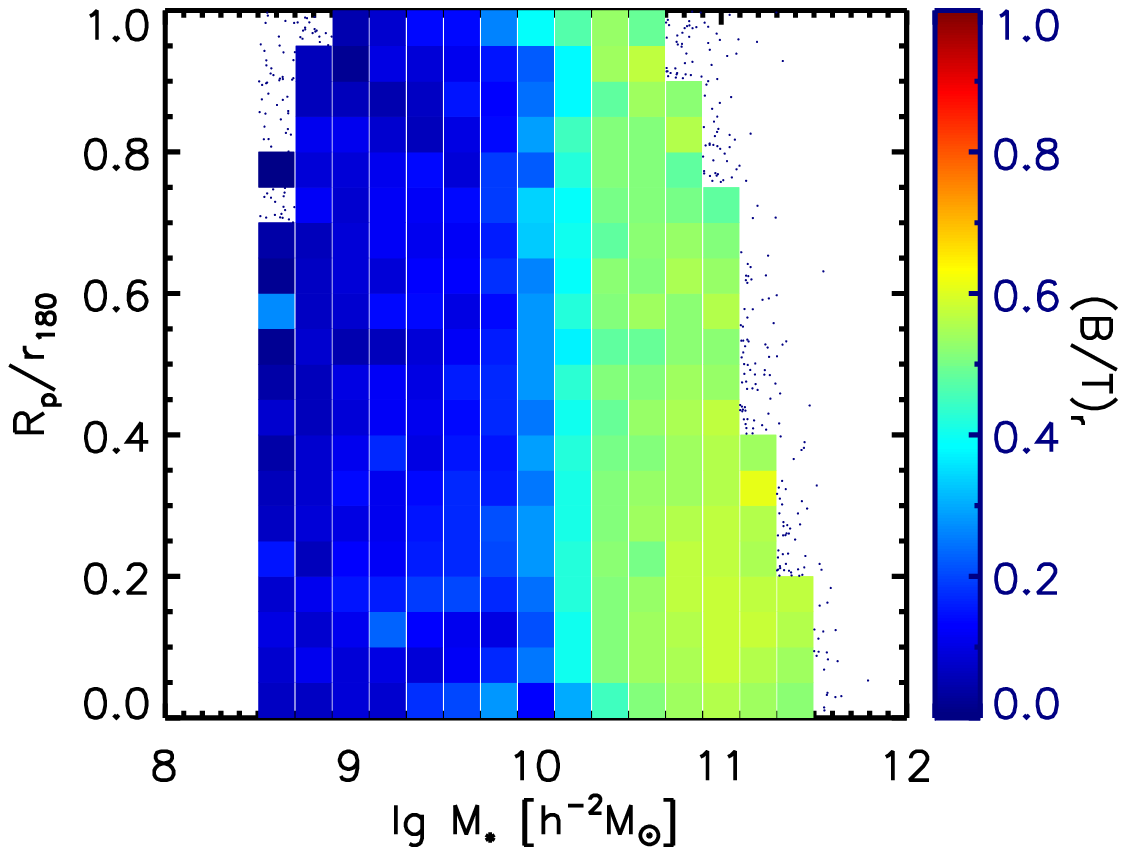}\\
\caption{Similar with Figure \ref{f_Mhs}, but on the $R_{p}/r_{180} - \lg M_*$ maps. $R_{p}/r_{180}$ is the halo centric radius. 
}
\label{f_RMs}
\end{figure*}

Finally, we check the dependence on another important halo-scale parameter -- the halo-centric radius $R_p/r_{180}$, which is the distance to the center of the halo of the group where the galaxy is located. We show in Figure \ref{f_RMs} the median Hubble types T and the median bulge-to-total light ratio of galaxies on a $R_{p}/r_{180} - \lg M_*$ map.  Bins with similar T and similar B/T are both distributed vertically. This suggests that the relative position to the center of the halos is also not important in the morphological transformation. These all suggest that the morphological transformation of galaxies is solely an in-situ process.

\subsection{The star formation quenching}
\label{sec_dis_quench}

\begin{figure*}[!htp]
\centering
\includegraphics[width=0.9\textwidth]{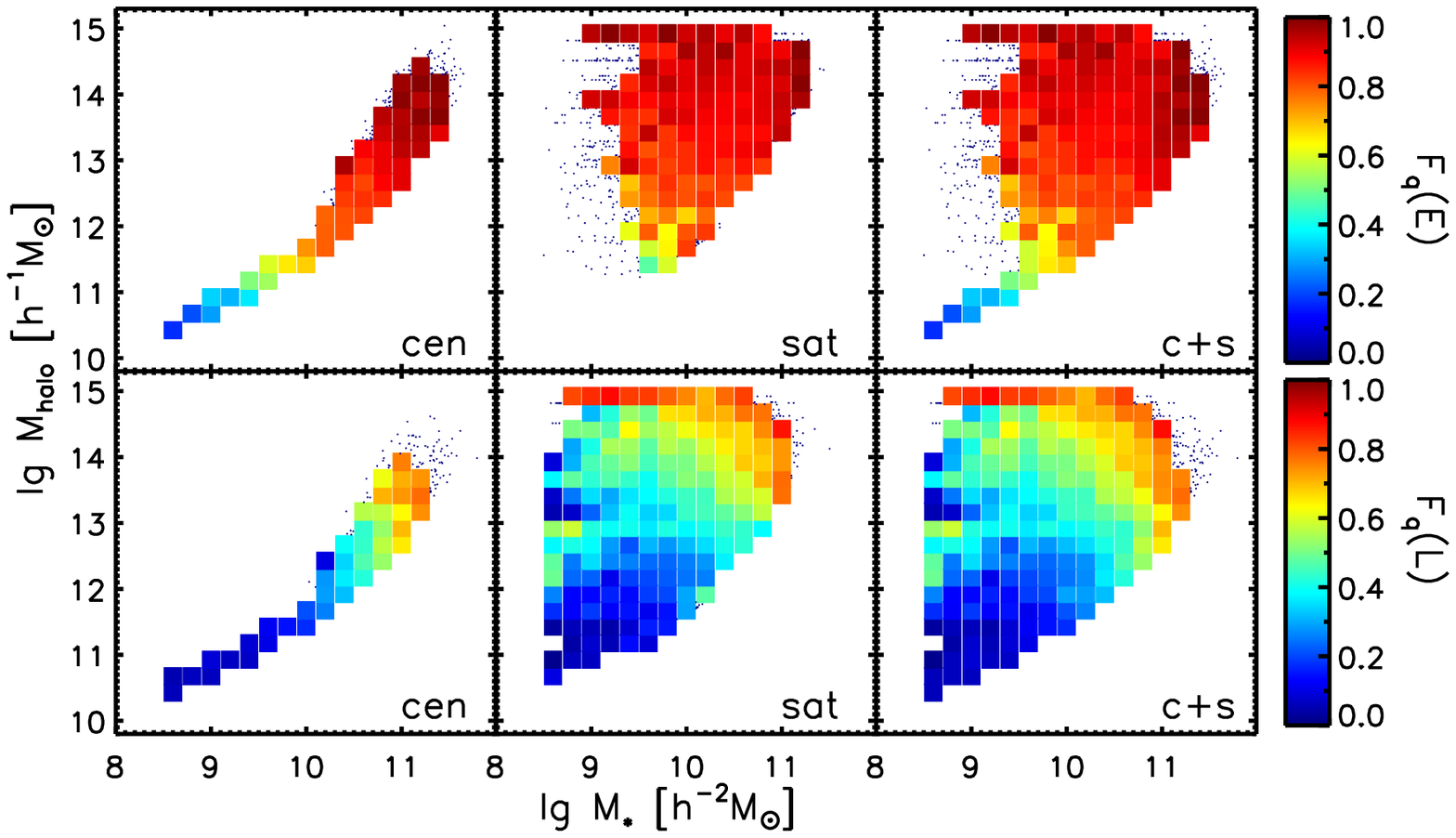}\\
\caption{Quenched fractions at fixed morphology on the $\lg M_{halo}-\lg M_*$ diagram. Bins are color coded by their quenched fractions, as are shown by the color bars. Upper panels are for the fractions of quenched early type galaxies out of all early type galaxies ($F_q(E)$). Lower panels are for the fractions of quenched late type galaxies out of all late type galaxies ($F_q(L)$). Left to right: centrals, satellites, and (cen$+$sat). Only bins with $n(E)\ge20$ for the upper panel and $n(L)\ge20$ are shown for the bottom panel. For the other bins that do not satisfy the above criteria, we show the positions of the quenched galaxies as black dots.
}
\label{f_Fmorph}
\end{figure*}

In Figure \ref{f_Fmorph}, we plot the fraction of quenched galaxies among the early type galaxies (upper panel) and among the late type galaxies (lower panel). Again, 
the $M_{*}$ and the $M_{halo}$ for centrals show tight correlation, and it is hard to tell the dominant parameter to control the changes of the quenched fraction.
For satellites (middle panel), the color changes more dramatically in the vertical direction in more massive halos ($M_{halo}\lesssim10^{13}\ M_{\sun}$) and more dramatically in the horizontal direction at higher stellar masses ($M_{*}\gtrsim10^{10.5}\ M_{\sun}$). Our results suggest that the quenching of massive galaxies is more of an in-situ process, while the quenching of satellites with lower stellar masses in halos with $M_{halo}\ga 10^{13.0}\msunh$ are dominated by their halo environments.\\

\begin{figure*}
\centering
\includegraphics[width=0.45\textwidth]{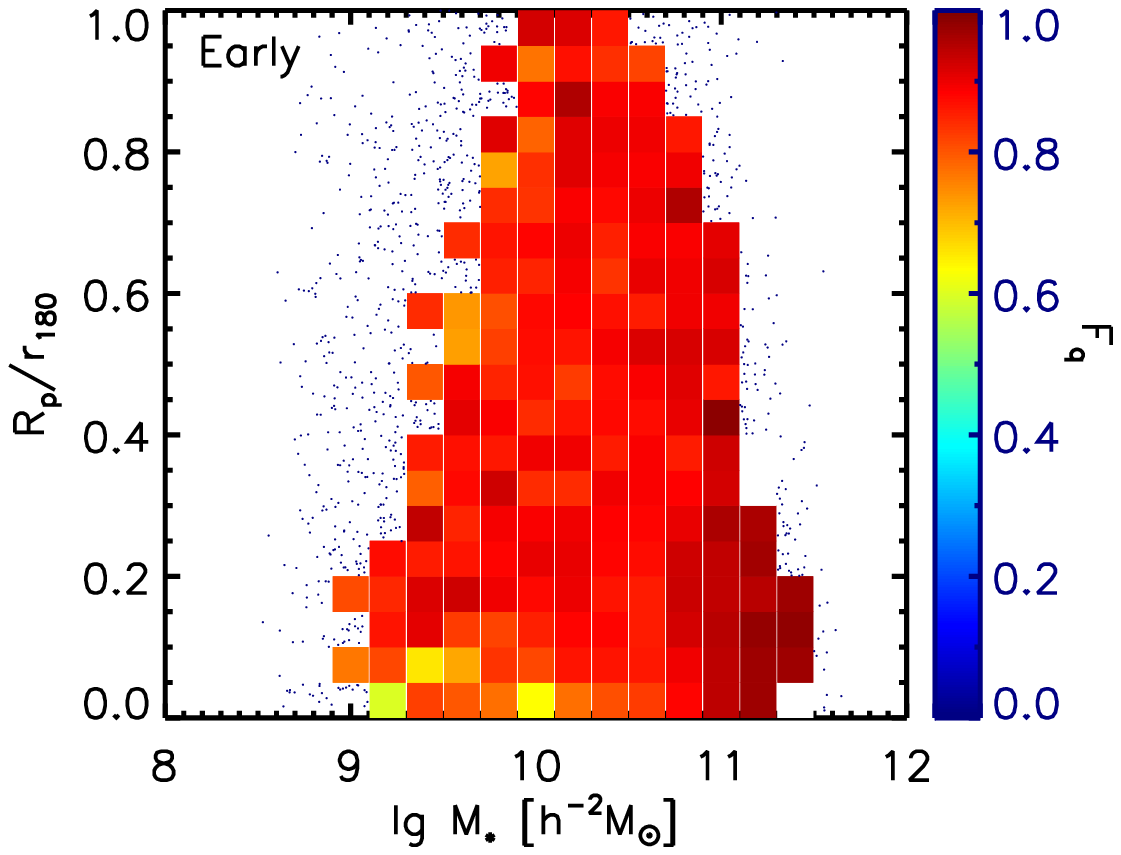}
\includegraphics[width=0.45\textwidth]{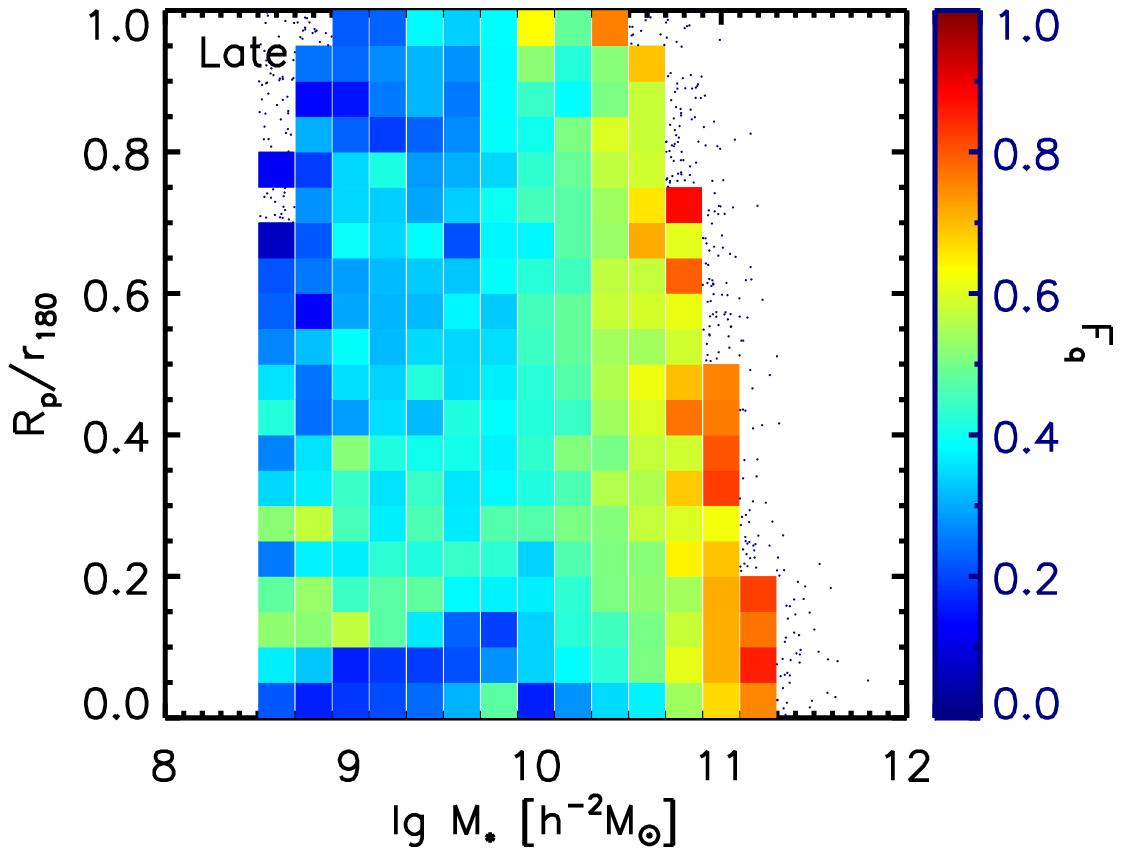}
\caption{Left: The quenched fraction among early type galaxies on the $R_{p}/r_{180} - \lg M_*$ map. Right: Same with the left panel, but for the quenched fraction among late type galaxies. Only bins with $n(gal)\ge20$ are shown. For bins with $n(gal)<20$, galaxies are shown as black dots. \\
}
\label{f_MRhs_Fq}
\end{figure*}

We show the distribution of the fraction of the quenched population ($F_q$) on the $R_{p}/r_{180} - \lg M_*$ map 
for the early types (left) and the late types (right) in Figure \ref{f_MRhs_Fq}.  
Bins with similar color coding are mainly vertically distributed and slightly horizontally distributed for galaxies with small distances to the center of the halo ($R_{p}/r_{180}<0.2$). 

Our result that quenching is mainly controlled by the stellar mass at high stellar masses, and by the halo mass 
at low stellar masses (Figure \ref{f_small_Fq} and Figure \ref{f_Fmorph}) agrees with the model raised by \cite{Peng2010}. They established a simple model that mass quenching dominates in the more massive ranges ($M_* > 10^{10.6}\ M_{\sun}$ at z=0), and environmental quenching dominates in the less massive ranges ($M_* < 10^{10.6}\ M_{\sun}$ at z=0). \cite{WangH2018} further found that the halo mass is the prime environmental parameter for quenching. Our results are consistent with their result that the quenching for galaxies with $M_* \lesssim 10^{10.5}\ M_{\sun}$ is probably a halo-scale physical process instead of those on larger environmental scales. The high quenching ratio in denser environments for the less massive galaxies are mostly caused by the less massive satellites in massive halos (Figure \ref{f_Fmorph}, middle panel). 

\cite{Peng2010} focused on the quenching of SF in their paper. 
Our paper supplies the supplemental morphology information to their study. 
Comparing the upper panel with the lower panel of Figure \ref{f_Fmorph}, we see an overall more significant halo quenching effect in early type galaxies than in late type galaxies. This might be attributed to their lacking of cold gas or earlier accretion into the massive host halos.

\subsection{
The morphology-density relation and the color-density relation}\label{sec_dis_relations}

From our result, we think the observed morphology-density relation may actually originate from the different $M_*$ distributions in environments with different densities. Denser environments are more abundant in more massive galaxies and less dense environments are more abundant in less massive galaxies \citep[e.g.][]{Constantin2008,Hoyle2012,Liu2015}. 
As shown in our paper, morphology transformation is more regulated by the stellar mass, 
therefore, denser environments statistically include more early type galaxies and less dense environments are of more later type galaxies. These observed morphology-density relation may intrinsically just be the $M_*$-density relation and the morphology-$M_*$ relation. We show that once the stellar mass is fixed, the morphology-density relation disappears (the top panel of Figure \ref{f_small_FE}).\\

Similarly, the color-density relation (or the SFR-density relation) originates from the $M_*$-density relation and the $M_{halo}$-density relation. We show that when the $M_*$ is fixed at high stellar masses and when the $M_{halo}$ is fixed at the low mass end, the color-density relation/SFR-density relation disappears (the top panel and the middle panel of Figure \ref{f_small_Fq}). \\

\cite{Bamford2009} studied the Galaxy Zoo sample and found no significant relation between the morphology evolution and the environments with the color fixed, while at fixed morphologies, the environmental dependence of color is strong, particularly for lower stellar-mass galaxies.  They used the local density as the indicator of the environments, which has different implications from our large-scale environment parameters. Similar findings are also mentioned in \cite{Skibba2009}, even though they probed environmental effect through the two-point clustering statistics. We have similar results, in particular, we also find that at lower stellar mass, the quenched fractions (red color) among fixed morphology strongly correlate with the large-scale environment (Figure \ref{f_small_Fq} upper left).  However, these variations disappear when the halo masses are fixed (Figure \ref{f_small_Fq} middle left). More detailed comparison of our conclusion and \cite{Bamford2009} and \cite{Skibba2009} can be complicated, as the color and the SFR are not strictly correlated \citep{Woo2013}, and the local density is sometimes difficult to interpret \citep[e.g.][]{Woo2013,WangH2018}

\section{conclusions}
\label{sec_conclusions}

We study the quenching and morphological transformation of galaxies with the seventh Data Release (DR7) of the Sloan Digital Sky Survey (SDSS). Both early type galaxies and the late type galaxies (as identified using a machine learning algorithm trained on expert classifications of local galaxies, see more details in Section \ref{sec_data_morph}) are seen to have bi-modal distributions on the $\lg SFR - \lg M_*$ diagram. We therefore classify them into four types: the star-forming early types (sEs), the quenched early types (qEs), the star-forming late types (sLs) and the quenched late types (qLs). Their abundances are $f_{sE}=4.1\%$, $f_{qE}=11.6\%$, $f_{sL}=67.2\%$, and $f_{qL}=17\%$. 

We checked many parameters that are related with physical processes on various scales (
the stellar mass $M_*$; on small scales, the halo mass $M_{halo}$, the halo centric radius $R_p/r_{180}$, and the third nearest neighbor distances $d_{3nn}$; on large scales, the cosmic web structures) for their potential effects on the morphological transformation and the quenching of SF. We found that 
the morphological transformation is mainly regulated by the stellar mass. Quenching is mainly driven by the stellar mass for more massive galaxies and by the halo mass for galaxies with smaller stellar masses.

Once the stellar mass is fixed, the morphology-density relation disappears. Similarly, once the stellar mass is fixed at the high mass end and the halo mass is fixed at the low mass end, the SFR-density relation disappears as well.

At small stellar masses ($M_* \lesssim 10^{10.5}\ \rm h^{-2}M_{\sun}$), quenching is dominated by the halo mass (Figure \ref{f_Fmorph}), while the morphology is dominated by the stellar mass (Figure \ref{f_Fqs}). This may suggest that the quenching of star formation and the morphological transformation could be two independent processes for less massive galaxies. 

At high stellar masses ($M_* \gtrsim 10^{10.5}\ \rm h^{-2}M_{\sun}$), both the quenching of star formation and the morphological transformation is mainly regulated by the stellar mass (Figure \ref{f_Fmorph} and Figure \ref{f_Fqs}). Figure \ref{f_Fmorph} also shows that the quenching fraction of the early type galaxies is higher than that of the late type galaxies, which may be the result of their lacking of cold gas or earlier accretion into the massive host halos. These may suggest that the quenching of star formation and the morphological transformation could be correlated for massive galaxies.


\acknowledgments
\noindent {\bf Acknowledgments:}
\vspace{0.2in}

This work is support by the National Natural Science Foundation of China (Nos. 11473055, 11733004, 11421303, 11522324, 11621303, 11833005, U1831205, 11603058) and the National Key R\&D Program of China (grant No. 2015CB857002). C.L. acknowledge support from China Scholarship Council (CSC). 



\appendix
\section[]{The redshift bias}
\label{sec:AppA}

In this section, we checked whether our main results -- ``The morphological transformation is mainly regulated by the stellar mass, and quenching is mainly driven by the stellar mass for more massive galaxies and by the halo mass for galaxies with smaller stellar masses'' are affected by redshift bias. 

\begin{figure*}[!htp]
\centering
\includegraphics[width=0.48\textwidth]{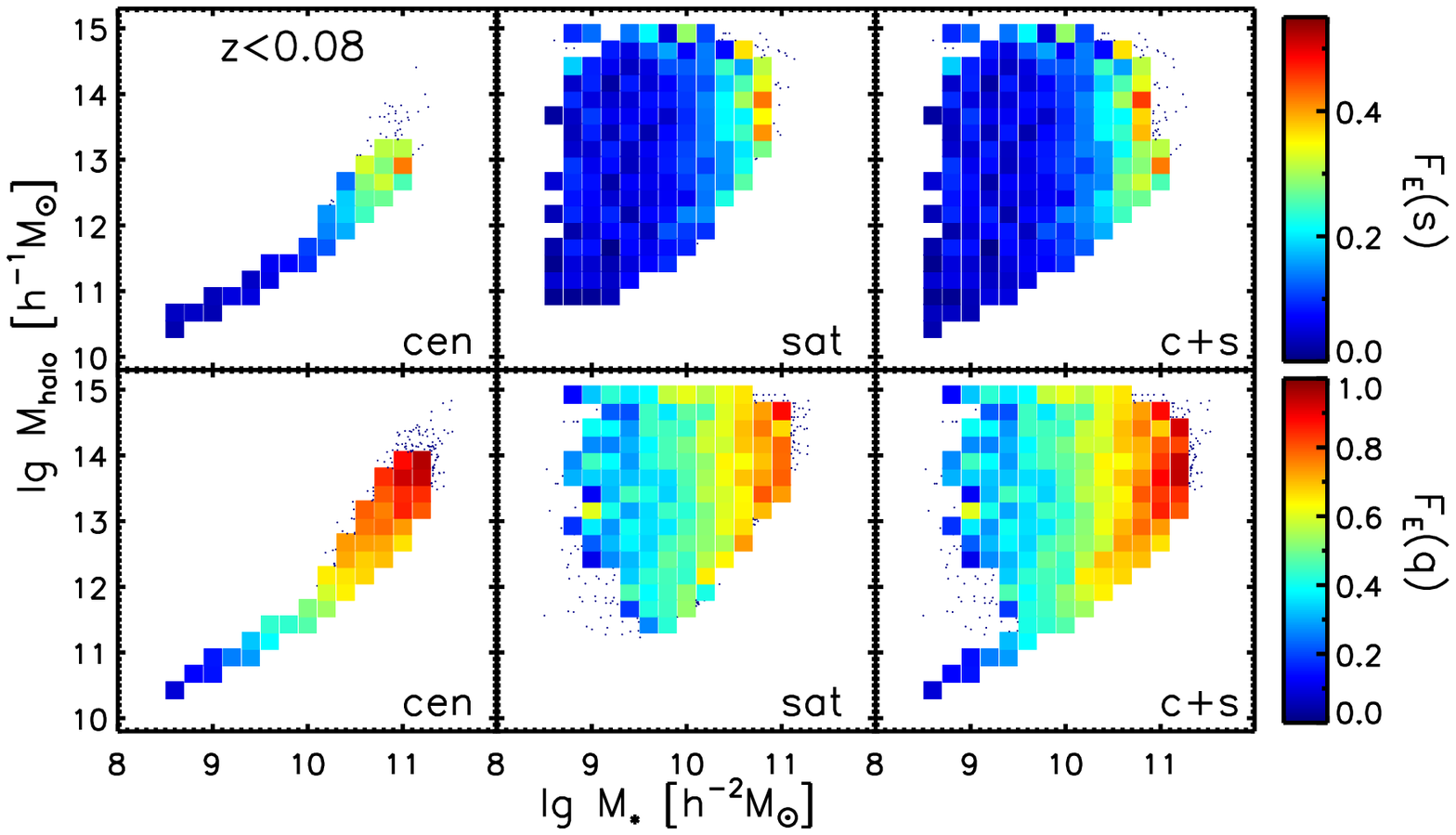}
\includegraphics[width=0.48\textwidth]{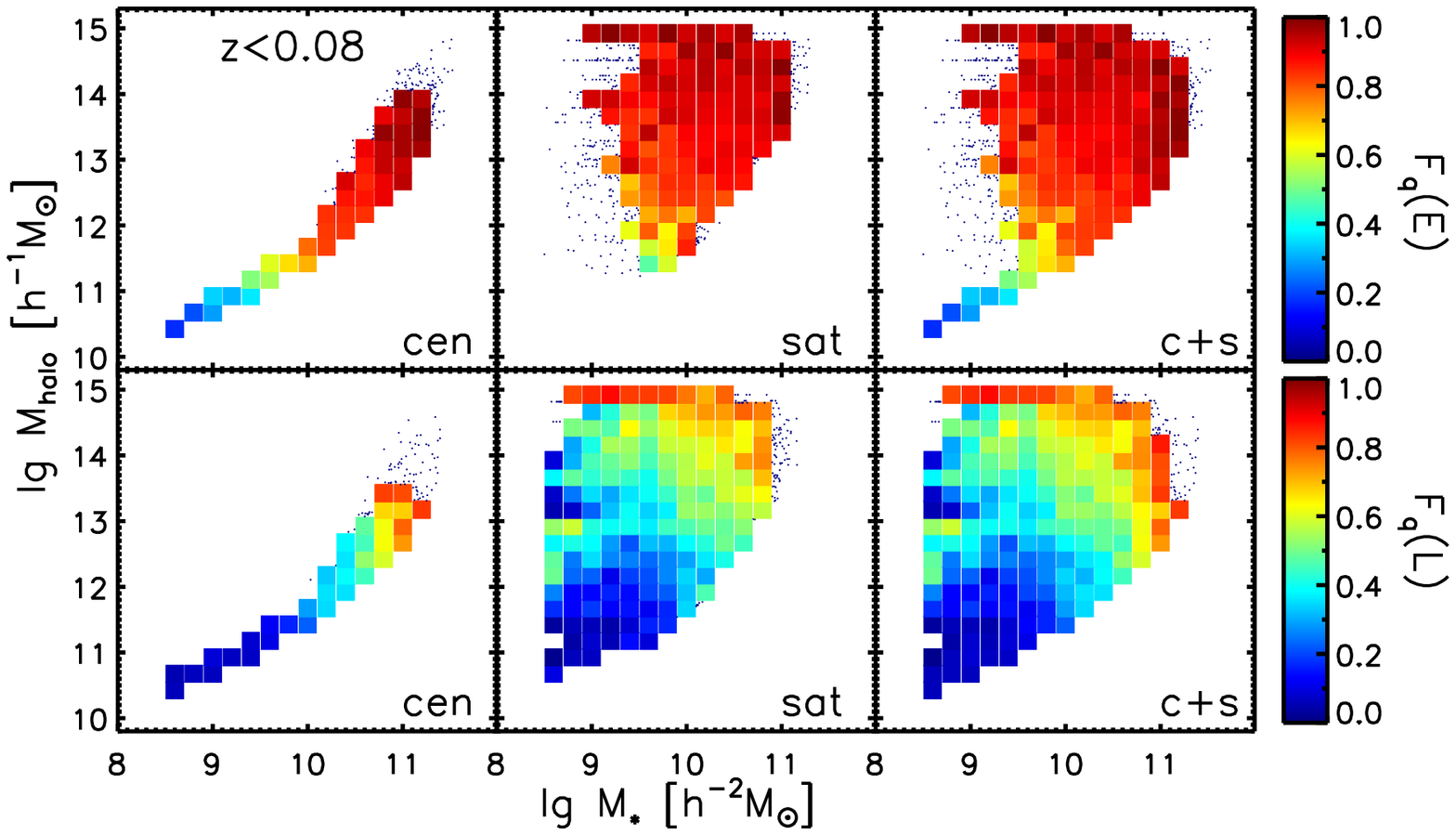}\\
\caption{Similar plots with Figure \ref{f_Fqs} and Figure \ref{f_Fmorph}, but with a lower redshift cut at $z<0.08$.
}
\label{f_lowz}
\end{figure*}

We first apply the redshift cut at $z<0.08$ and our main results qualitatively hold, as is shown in Figure \ref{f_lowz}. We also checked lower redshift cuts, such as $z<0.05$, our main results do not change.

\begin{figure}[!htp]
\centering
\includegraphics[width=0.45\textwidth]{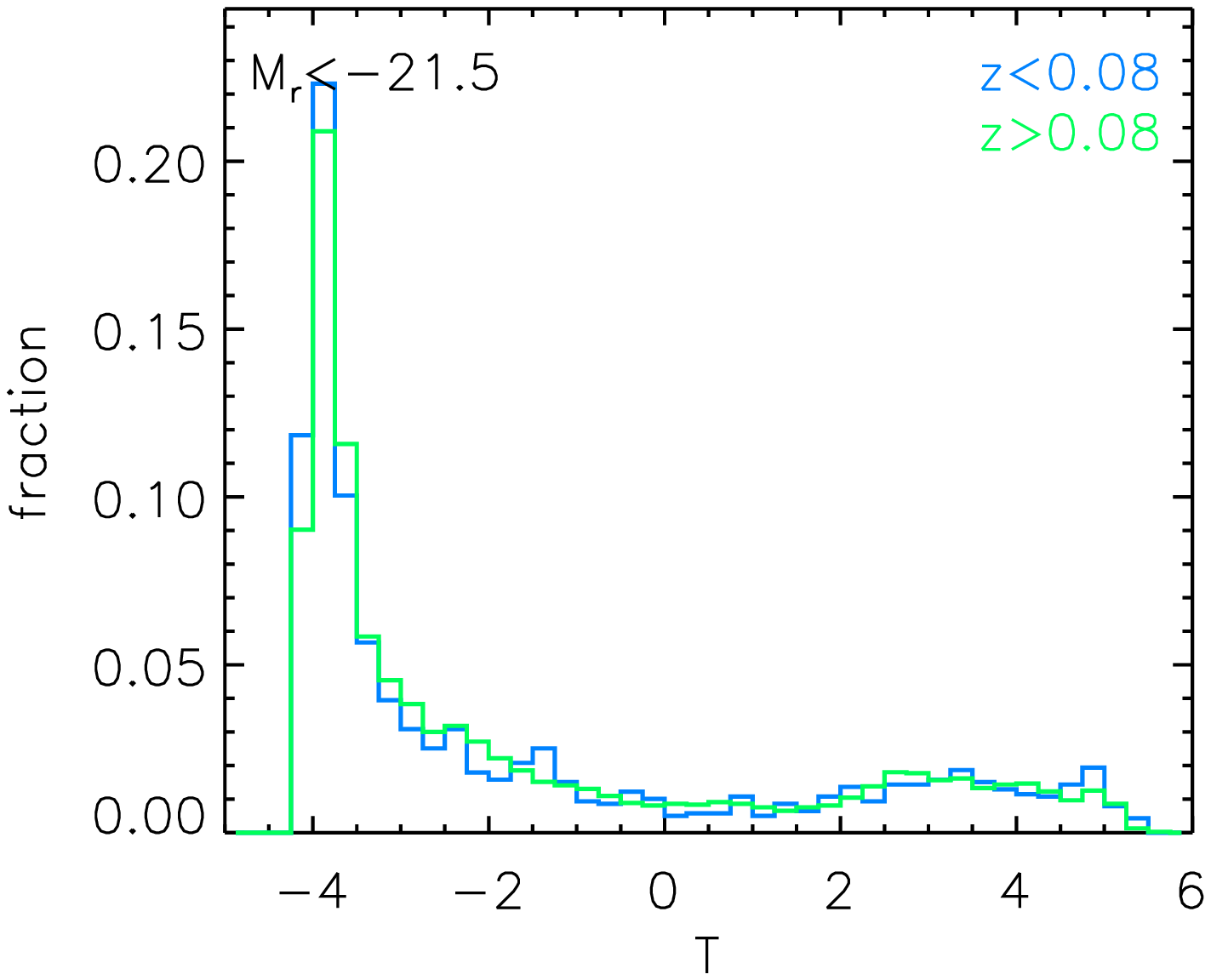}
\includegraphics[width=0.45\textwidth]{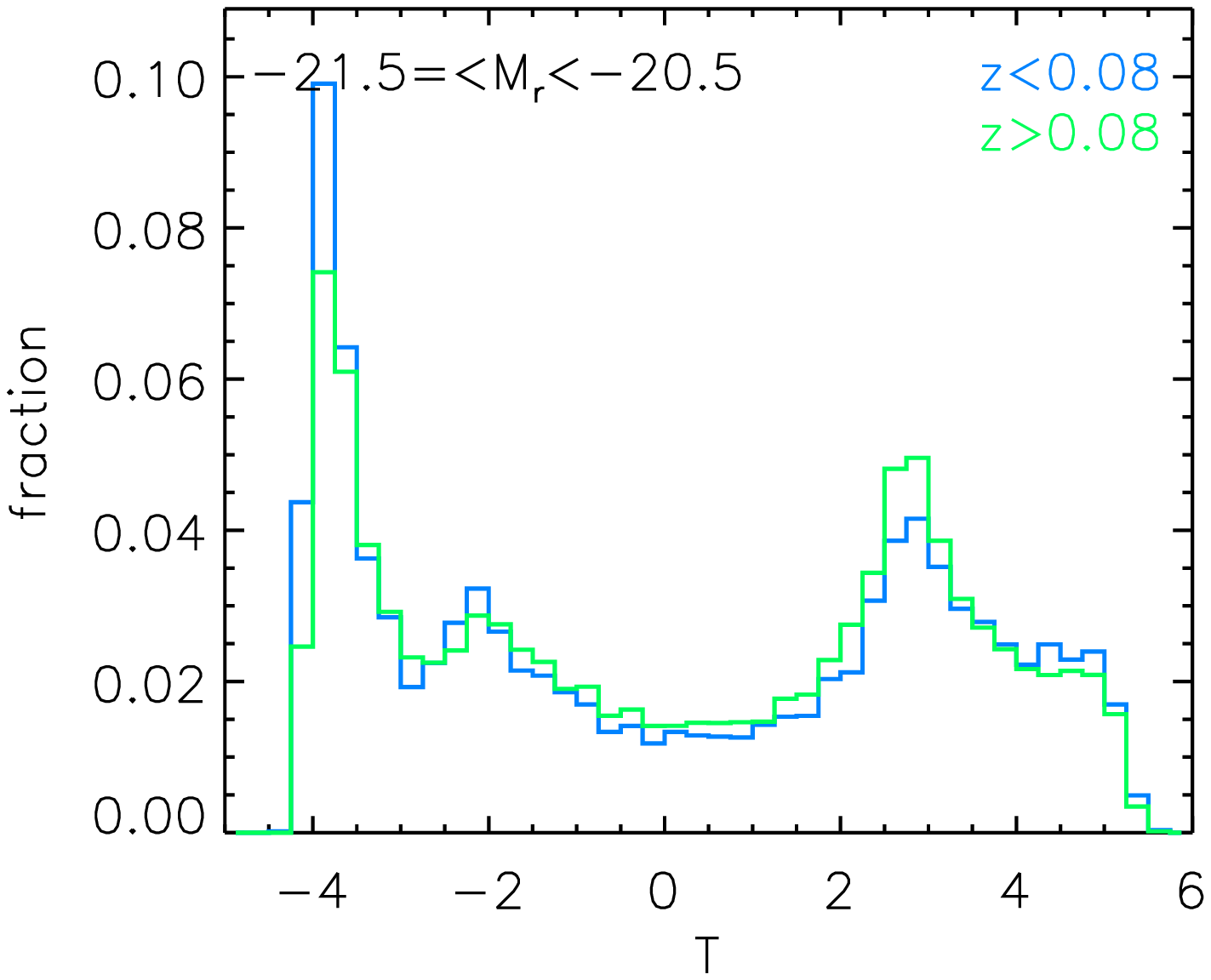}\\
\includegraphics[width=0.45\textwidth]{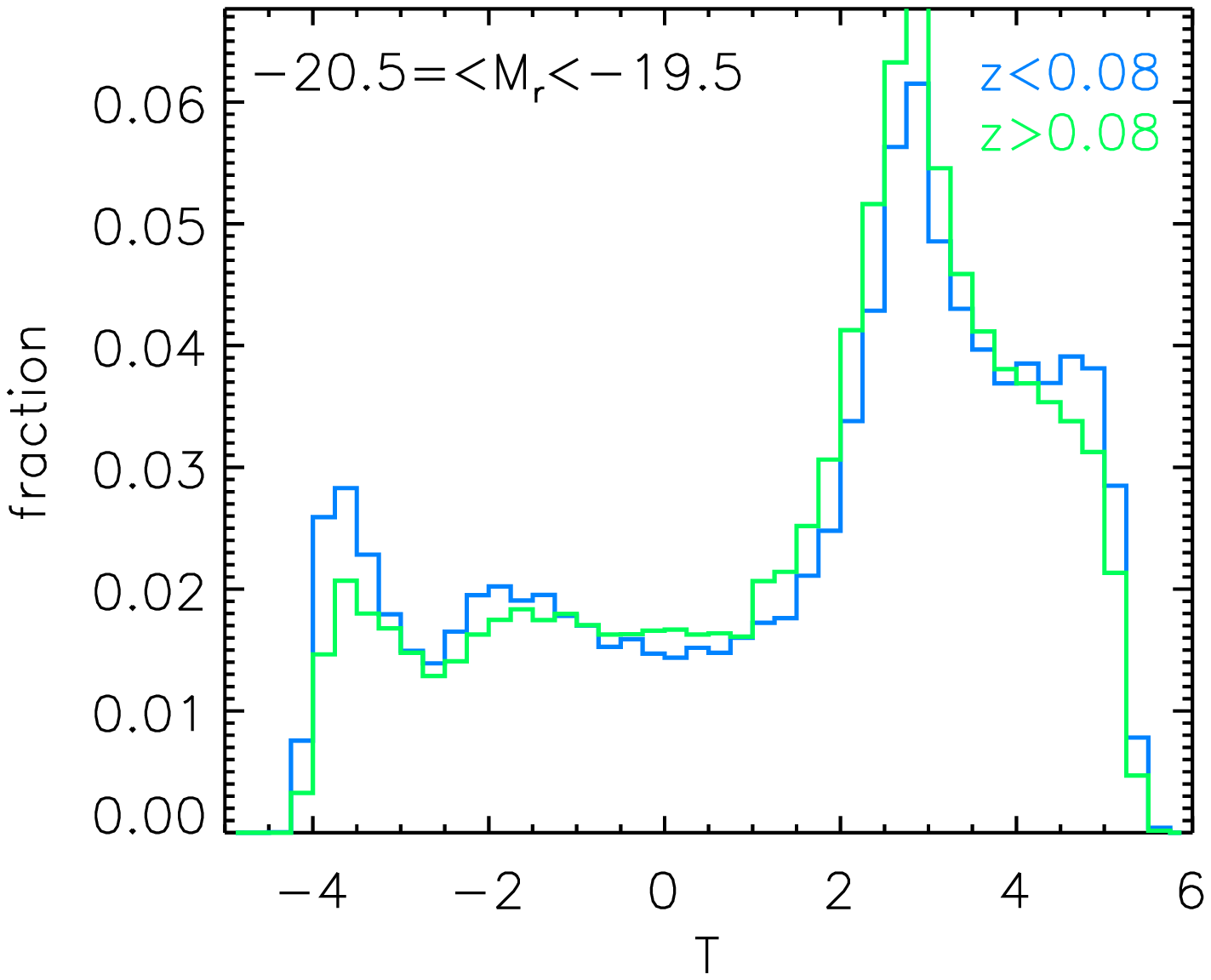}
\includegraphics[width=0.45\textwidth]{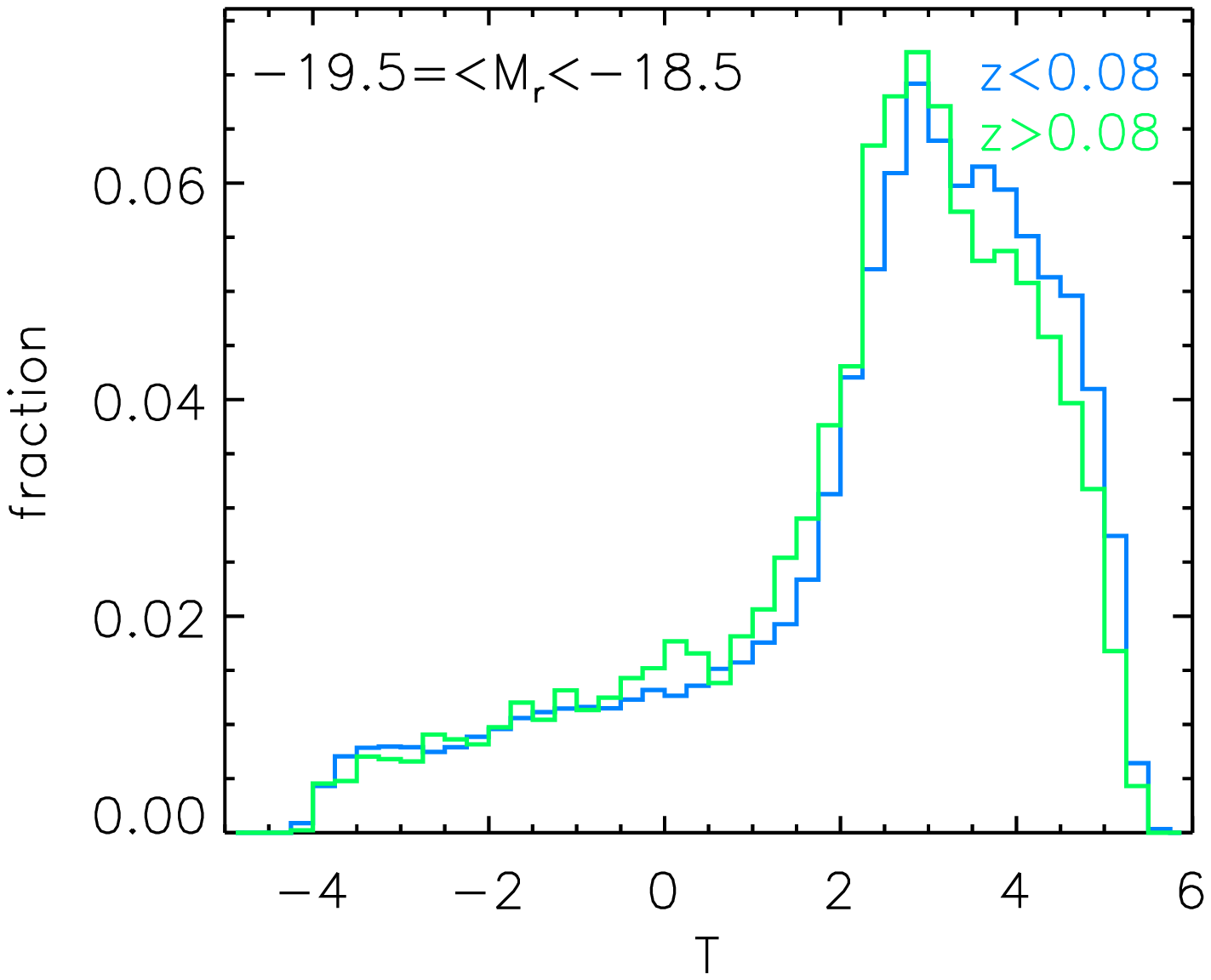}\\
\caption{The Hubble type T distributions for the lower redshift $z<0.08$ subsample and the higher redshift $z>0.08$ subsample in different luminosity bins.}
\label{f_Tdistr}
\end{figure}

The redshift bias could come from the mis-idenfication of morphology types. \cite{Bamford2009} found that galaxies beyond $z=0.08$ are more likely to be identified as ellipticals due to resolution and surface brightness effects. The redshift can also affect the result with the SFR measurements, since the lower redshift galaxies would suffer more from the aperture correction uncertainties. We therefore checked the distributions of Hubble type T values for the $z>0.08$ subsample and the $z<0.08$ subsample in different luminosity bins in Figure \ref{f_Tdistr}. We didn't find significant differences in the T distributions between the high redshift $z>0.08$ subsample and the low redshift $z<0.08$ subsample.

\begin{figure}[!htp]
\centering
\includegraphics[width=\textwidth]{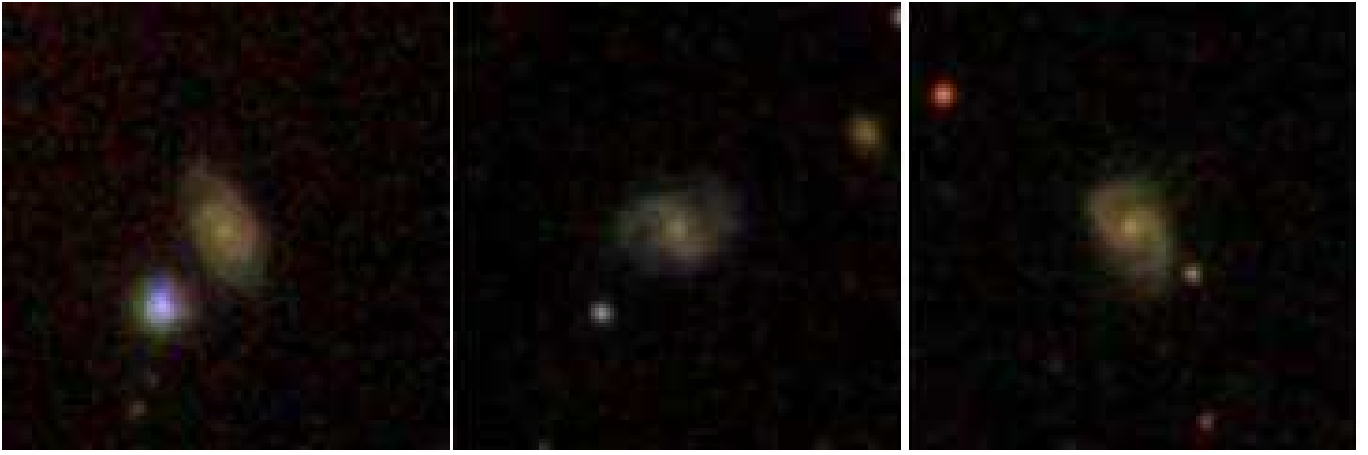}\\
\caption{Three random images late type galaxies at $z\approx0.12$. 
}
\label{f_images}
\end{figure}

We randomly checked some images of late type galaxies going out to $z\approx0.12$ in Figure \ref{f_images}. 
We can see that even the image of late types at $z>0.11$ does not allow us to see the morphologies with a lot of details, we can still recognize the spiral feature. 

The large aperture correction error for the SFR measurement of the low redshift galaxies. However, as we mentioned in Section \ref{sec_data_sfr}, our results rely more on the bi-modal feature in the SFR-$M_*$ diagram, instead of the absolute SFR values, because we care more about the quenched fraction. Our $z<0.08$ low redshift sample also indirectly shows that our results are not sensitive to redshift.

\end{document}